\documentclass[12pt]{iopart}
\usepackage{url}
\usepackage{multirow}
\usepackage{graphicx}
\usepackage{iopams}
\usepackage{hhline}
\usepackage{epsfig}
\usepackage{epstopdf}
\usepackage{booktabs}
\usepackage{multirow}
\usepackage{siunitx}
\usepackage{adjustbox}
\usepackage{float}
\usepackage{caption}
\usepackage{array}
\usepackage{cite}
\usepackage{appendix}
\usepackage{etoolbox}
\usepackage{graphicx, nicefrac}
\newcolumntype{L}[1]{>{\raggedright\let\newline\\\arraybackslash\hspace{0pt}}m{#1}}
\newcolumntype{C}[1]{>{\centering\let\newline\\\arraybackslash\hspace{0pt}}m{#1}}
\newcolumntype{R}[1]{>{\raggedleft\let\newline\\\arraybackslash\hspace{0pt}}m{#1}}

\begin{document}
\title[Reliability of the Brink-Axel Hypothesis]
{Impact of the Brink-Axel Hypothesis on Unique First-Forbidden $\beta$-transitions for $r$-process nuclei}

\author{Fakeha Farooq$^1$, Jameel-Un Nabi$^{2}$, Ramoona Shehzadi$^1$}
\address{$^1$ Department of Physics, University of the Punjab, Quaid-e-Azam Campus 54000, 
Lahore, Pakistan.}
\address{$^2$ University of Wah, Quaid Avenue, Wah Cantt 47040, Punjab, Pakistan.}

\ead{ramoona.physics@pu.edu.pk}

\begin{abstract}
Key nuclear inputs for the astrophysical $r$-process simulations are the weak interaction rates.  Consequently, the 
accuracy of these inputs directly affects the reliability of nucleosynthesis modeling. Majority of the stellar rates,  
used in simulation studies, are calculated invoking the Brink-Axel (\textit{BA}) hypothesis.  The \textit{BA} hypothesis assumes that the strength functions of all parent excited 
states are the same as for the ground state, only shifted in energies. However, \textit{BA} hypothesis has to be tested against microscopically  calculated 
state-by-state rates. In this project we study the impact of the \textit{BA} hypothesis on calculated stellar 
$\beta^{-}$-decay  and electron capture rates. Our investigation include both Unique First Forbidden (U1F) and allowed transitions  
for 106 neutron-rich trans-iron nuclei ([27, 77]\;$\leq$\;[Z, A]\;$\leq$\;[82, 208]).  
The calculations were performed using the deformed  proton-neutron  quasiparticle random-phase approximation (pn-QRPA) model with a simple plus quadrupole separable and schematic interaction.  Waiting-point and several 
key $r$-process nuclei lie within the considered mass region of the nuclear chart. We computed electron capture and $\beta^{-}$-decay rates using two different 
prescriptions for strength functions. One was based by invoking \textit{BA} hypothesis and the other was the state-by-state calculation of 
strength functions, under  stellar density  and temperature
conditions ([10,  1]\;$\leq$\;[$\rho\text{Y}_{e}$($g/cm^{3}$),  T($GK$)]\;$\leq$\;[10$^{11}$, 30]). 
Our results show that \textit{BA} hypothesis invoked U1F $\beta^{-}$ rates are overestimated by 4--5 orders of magnitude as compared to microscopic rates. For capture rates, more than 2  orders of magnitude difference  was noted when applying  \textit{BA} hypothesis. 
It 
was concluded that the \textit{BA} hypothesis is not a reliable approximation, especially for the $\beta^{-}$-decay  forbidden transitions.         

\end{abstract}


\maketitle
\section{Introduction}
\label{sec:intr}
In astrophysical measures, forbidden transitions are the key elements to study the nuclear weak processes 
($\beta^{\pm}$-decays and electron captures), in addition to the allowed transitions~\cite{Hom96,Mol03,Zhi13,Nab14}. 
When the electron chemical potential of the stellar interior approaches $\sim$30 [MeV], 
first-forbidden (FF) transitions become important relative to the allowed Gamow-Teller (GT) rates. 
The $\beta^{-}$-decay half-lives of nuclei having  large Z number and with Z and N beyond closed shells have 
sizable contribution from FF decays~\cite{Suz11, Nab15}. Reliable calculation of weak rates, including $\beta^{-}$-decay (BD) and 
electron capture (EC), on heavy trans-iron neutron-rich nuclei (70\;$\leq$\;A\;$\leq$\;208) are a prerequisite to study the rapid neutron-capture ($r$-) process~\cite{Nab04,Cla68,Ney20}. 
Particularly, the weak rates associated with FF BD transitions on closed-shell waiting-point (WP) nuclei, with N numbers 50, 82 and 126, 
have potential contributions in affecting the progression of matter in the $r$-process pathway. The observed peaks in the abundance pattern of $r$-process elements arise due to the deceleration of matter flow at these WPs.  Thus, matter assembles at the WP and nuclei undergo a series of BD before the $r$-process recommences~\cite{Cow21,Hor19,Nabi19,Nabi2021,Nabi21,Mol03,Mar16,Bor0811}. The high temperature ($>$ 1 [GK]) and 
high neutron density ($>$ 10$^{20}$ [g/cm$^{3}$]) conditions associated with the neutron-star to neutron-star 
collisions~\cite{Wat19} and core-collapse supernovae (CCSNe)\cite{Hor19} establish a site for the creation of the 
$r$-process elements. 

The weak rates are calculated on the basis of charge-changing strength functions or transition probabilities. The charge-exchange 
reactions provide GT strength distribution data from the ground states of high mass nuclei (e.g., ~\cite{Thi12,Fre18,Pup12,Yas18,Gue11,Wak12}). 
However,  measurements can provide information only for a limited number of nuclei. In addition, in exotic 
stellar environment, states can be thermally populated, and transitions from the excited states, which, in general, are 
not measurable under lab conditions, contribute significantly\cite{Nab04,Lan00}. Despite the improved measurement facilities on new heavy-ion accelerators 
(e.g.,~\cite{Nis11,Lor15,Dom19,Liang20,Hal21}), it seems a daunting challenge  to obtain satisfactory and accurate 
information from high-lying excited states using experiments. Consequently, astrophysical simulations rely heavily on  
theoretical estimations. However, for nuclear models, it is still a challenging job to perform rate calculations near neutron (proton) drip line and region of higher mass nuclei, due to the complex structures 
of correlated many-body nuclear systems. Majority of the previously calculated set of weak rate calculations\cite{Fuller,Auf94,Lan03,Lan00,Col12} used an approximate 
method invoking the so-called Brink-Axel (\textit{BA}) hypothesis\cite{BriA}. In several studies, the accuracy of the \textit{BA} hypothesis 
has been investigated~\cite{Fra97,Lar07,Nab07,Ang12,Mis14,Joh15,Lu18,Nab19,Yuk20,Her22,Nab22,Fak23}. Majority of these works have shown the fact that \textit{BA} hypothesis is a poor approximation for usage in calculations related to stellar weak rates. The test of the effectiveness of the \textit{BA} hypothesis is crucial for reliable 
estimation of stellar weak rates. To the best of our knowledge, earlier studies did not cover the $r$-process 
nuclei as well as the validity of the \textit{BA} hypothesis for computation of forbidden transitions. The present investigation reports the 
effectiveness of the \textit{BA} hypothesis for the calculations of stellar (EC and BD) weak rate. We selected  106 $r$-process 
nuclei (27\;$\leq$\;Z\;$\leq$\;82 and 70\;$\leq$\;A\;$\leq$\;208) and studied their allowed and U1F transitions under stellar conditions. The calculation of terrestrial BD half-lives and $\beta$-delayed neutron-emission probabilities of the nuclei has remained the focus 
of many past investigations. These include experimental~\cite{Nis11,Lor15,Dom19,Hal21,Pfe02} as well as theoretical  studies based on shell model~\cite{Lan03,Cue07,Zhi13}, 
QRPA +Gross theory~\cite{Mol03},   density functional + continuum QRPA~\cite{Bor0811,Bor0305}, empirical calculations~\cite{Zho17}, machine learning~\cite{Niu19}, relativistic Hatree-Bogoliubov + QRPA
~\cite{Mar16}, pn-QRPA~\cite{Hir93, Sta90} and relativistic QRPA~\cite{Rob22}. 

The current manuscript is structured as follows. Section~\ref{sec:formalism} briefly describes the pn-QRPA formalism  for 
the calculation of strength functions and decay rates for both U1F and allowed (GT) transitions. Results 
are discussed in Section~\ref{sec:results}. Finally, Section~\ref{sec:conclusions} highlights the  findings of current investigation. 

\section{Theoretical Formalism}
\label{sec:formalism}
A simple and microscopic theoretical frame-work based on the pn-QRPA theory was employed in order to check the reliability of 
\textit{BA} hypothesis for weak rate calculations including U1F and allowed GT transitions. By using a separable and schematic interaction, access up to 7 major oscillatory shells model space became possible in the current calculation. This enabled us to calculate strength functions 
in a state-by-state fashion for high-mass nuclei considered in this project. This simple yet effective microscopic 
approach has wide applications in astrophysical studies\cite{Nab14,Nab15,Nab16,Far21}.

The pn-QRPA theory deals with quasiparticle states of proton-neutron systems and correlations between them. 
The ground state is a vacuum for QRPA phonon, $\hat{\Gamma}_{\omega}|$QRPA$> = 0$, with phonon creation operator defined by
\begin{eqnarray}
\hat{\Gamma}^{\dagger}_{\omega}(\mu)&=& \sum_{\pi,\nu}X^{\pi\nu}_{\omega}(\mu)\hat{a}^{\dagger}_{\pi}\hat{a}^{\dagger}_{\bar{\nu}}-Y^{\pi\nu}_{\omega}(\mu)\hat{a}_{\nu}\hat{a}_{\bar{\pi}},
\label{Eq:PCO}
\end{eqnarray}
where $\nu$ and $\pi$, respectively, denote the neutron and proton single quasiparticle states, $(\hat{a}^{\dagger}, \hat{a})$ are creation and annihilation operators of these states. The sum runs over all possible
$\pi\nu$-pairs which satisfy $\mu=m_{\pi}-m_{\nu}$ with $m_{\pi}(m_{\nu})$ being the third component of angular momentum.
The forward-going ($X_{\omega}$) and backward-going ($Y_{\omega}$) amplitudes, and energy ($\omega$) are the eigenvectors and eigenvalues, respectively, of the
famous (Q)RPA equation
\begin{center}
\begin{equation}
  \left[ {\begin{array}{cc}
    M & N \\
  -N & -M \\
  \end{array} } \right]
   \left[ {\begin{array}{c}
    X \\
    Y\\
  \end{array} } \right] =  \omega \left[ {\begin{array}{c}
    X\\
    Y\\
  \end{array} } \right].
  \label{Eq:RPAM}
\end{equation}
\end{center}

The solution of RPA equation (Eq.~\ref{Eq:RPAM}) was obtained for each projection value ($\mu = 0,\pm1$ for allowed and $\mu = 0,\pm1,\pm2$ for U1F transitions). $M$ and $N$ matrix elements are given by
\begin{eqnarray}
M_{\pi\nu,\pi^{\prime}\nu^{\prime}}=&\delta_{\pi\nu,\pi^{\prime}\nu^{\prime}}(\varepsilon_{\pi}+\varepsilon_{\nu})\nonumber \\
&+V^{pp}_{\pi\nu,\pi^{\prime}\nu^{\prime}}(v_{\pi}v_{\nu}v_{\pi^{\prime}}v_{\nu^{\prime}}+u_{\pi}u_{\nu}u_{\pi^{\prime}}u_{\nu^{\prime}})\nonumber \\
&+V^{ph}_{\pi\nu,\pi^{\prime}\nu^{\prime}}(v_{\pi}u_{\nu}v_{\pi^{\prime}}u_{\nu^{\prime}}+u_{\pi}v_{\nu}u_{\pi^{\prime}}v_{\nu^{\prime}}),
\label{Mentry}
\end{eqnarray}
\begin{eqnarray}
N_{\pi\nu,\pi^{\prime}\nu^{\prime}}=&V^{pp}_{\pi\nu,\pi^{\prime}\nu^{\prime}}(u_{\pi}u_{\nu}v_{\pi^{\prime}}v_{\nu^{\prime}}+v_{\pi}v_{\nu}u_{\pi^{\prime}}u_{\nu^{\prime}})\nonumber \\
&-V^{ph}_{\pi\nu,\pi^{\prime}\nu^{\prime}}(v_{\pi}u_{\nu}u_{\pi^{\prime}}v_{\nu^{\prime}}+u_{\pi}v_{\nu}v_{\pi^{\prime}}u_{\nu^{\prime}}),
\label{Nentry}
\end{eqnarray}
where the quasiparticle energies ($\varepsilon_{\pi}, \varepsilon_{\nu}$) and the occupation
amplitudes ($u_{\pi(\nu)},v_{\pi(\nu)}$), which satisfy $u^{2}+v^{2} = 1$, were obtained from the BCS calculations.
In the first step, quasiparticle basis was constructed in terms of nucleon states and
defined by Bogoliubov transformation with pairing correlations.
Later, in the quasiparticle proton-neutron pairs,
the computation of the RPA equation (Eq.~\ref{Eq:RPAM}) was performed with separable GT residual forces, namely:
particle-hole (ph) and particle-particle (pp) forces. We took pp GT force as~\cite{Kuz88}
\begin{equation}
\hat{V}_{pp(GT)} = -2\kappa_{GT}\sum_{\mu} (-1)^{\mu}\hat{P}^{\dagger}_{\mu}\hat{P}_{-\mu},
\label{ppGT}
\end{equation}
where
\begin{equation}
\hat{P}^{\dagger}_{\mu} = \sum_{j_{\pi}m_{\pi}{j_{\nu}m_{\nu}}} \langle j_{\nu}m_{\nu}|(\sigma_{\mu}\tau_{-})^{\dagger}|j_{\pi}m_{\pi} \rangle (-1)^{l_{\nu}+j_{\nu}-m_{\nu}}\hat{c}^{\dagger}_{j_{\pi}m_{\pi}}\hat{c}^{\dagger}_{j_{\nu}-m_{\nu}},
\end{equation}
and ph GT force as~\cite{Halb67}
\begin{equation}
\hat{V}_{ph(GT)} = 2\chi_{GT}\sum_{\mu} (-1)^{\mu}\hat{R}_{\mu}\hat{R}^{\dagger}_{-\mu},
\label{phGT}
\end{equation}
where
\begin{equation}
\hat{R}_{\mu} = \sum_{j_{\pi}m_{\pi}{j_{\nu}m_{\nu}}} \langle j_{\pi}m_{\pi}|\sigma_{\mu}\tau_{-}|j_{\nu}m_{\nu} \rangle\hat{c}^{\dagger}_{j_{\pi}m_{\pi}}\hat{c}_{j_{\nu}m_{\nu}}.
\end{equation}
Introducing positive values of force constants ($\chi_{GT}, \kappa_{GT}$), ensured attractive and repulsive nature of pp and ph GT forces, respectively. With the use of the separable GT forces in the calculation, the RPA matrix equation reduces to a 4$^{th}$ order algebraic equation. The method to determine the roots of these equations can be seen from~\cite{Mut92}. This saves the computational time relative to the fully diagonalization of the
nuclear Hamiltonian.

In the RPA formalism, excitations from the ground state ($J^{\pi} = 0^+$) of an even-even nucleus is considered.
The ground-state of an odd-odd (odd-A) parent nucleus is expressed as a proton-neutron quasiparticle pair
(one quasiparticle) state of the smallest energy. Then two possible transitions are the phonon excitations (in which
quasiparticle merely plays the role of a spectator) and transition of quasiparticle itself. In
the latter case, correlations of phonon to the quasiparticle transitions were treated in first-order perturbation~\cite{Mol90,Ben88}. We next present quasiparticle transitions, construction of phonon-related multi-quasiparticle states (representing nuclear excited levels of even-even, odd-A and odd-odd nuclei) and formulae
of GT transitions within the current model using the recipe given by~\cite{Mut92}.
The phonon-correlated one quasiparticle states are defined by
\begin{eqnarray}
	|\pi_{corr}\rangle~=~a^\dagger_{\pi}|-\rangle +& \sum_{\nu, \omega}a^\dagger_{\nu}A^\dagger_{\omega}(\mu)|-\rangle \nonumber~\langle-|[a^\dagger_{\nu}A^\dagger_{\omega}(\mu)]^{\dagger}H_{31}a^\dagger_{\pi}|-\rangle \nonumber \\
	&\times E_{\pi}(\nu,\omega)
\label{opn1}
\end{eqnarray}
\begin{eqnarray}
	|\nu_{corr}\rangle~=~a^\dagger_{\nu}|-\rangle +& \sum_{\pi, \omega}a^\dagger_{\pi}A^\dagger_{\omega}(-\mu)|-\rangle \nonumber~\langle-|[a^\dagger_{\pi}A^\dagger_{\omega}(-\mu)]^{\dagger}H_{31}a^\dagger_{\nu}|-\rangle \nonumber \\
	&\times E_{\nu}(\pi,\omega)
	\label{opn2}
\end{eqnarray}
with
	\begin{equation}
		E_{a}(b,\omega)=\frac{1}{\epsilon_{a}-\epsilon_{b}-\omega}~~~~~~~a, b = \pi, \nu
		\label{opn3}
	\end{equation}
where the terms $E_{a}(b,\omega)$ can be modified to prevent the singularity in the transition amplitude
caused by the first-order perturbation of the odd-particle wave function. The first term in Eq.~(\ref{opn1}) and Eq.~(\ref{opn2}) denotes the proton (neutron) quasiparticle state, while the second term denotes RPA correlated
phonons admixed with quasiparticle phonon coupled Hamiltonian H$_{31}$, which was accomplished by Bogoliubov
transformation from separable pp and ph GT interaction forces. The summation applies to all phonon states and
neutron (proton) quasiparticle states, satisfying $m_{\pi}-m_{\nu}=\mu$ with $\pi_{\pi}\pi_{\nu}=1$. Calculation of the quasiparticle transition amplitudes for correlated states
can be seen from~\cite{Mut89}. The amplitudes of GT transitions in terms of separable forces are
\begin{eqnarray}
		<{\pi_{corr}}|\tau_-\sigma_{\mu}|{\nu_{corr}}>~=~ q^U_{\pi\nu}+ 2\chi_{GT} [q^U_{\pi\nu}\sum\limits_{\omega}(Z^{-2}_\omega E_\pi(\nu,\omega)+Z^{+2}_{\omega}E_\nu(\pi,\omega)) \nonumber\\
		-q^V_{\pi\nu}\sum\limits_{\omega}Z^-_{\omega}Z^+_{\omega}(E_\pi(\nu,\omega)+E_\nu(\pi,\omega))]
		+2\kappa_{GT}[q_{\pi\nu}\sum\limits_{\omega}(Z^-_{\omega}Z^{--}_{\omega}E_\pi(\nu,\omega)-Z^+_{\omega}Z^{++}_{\omega}E_\nu(\pi,\omega)) \nonumber\\
		-\tilde{q}_{\pi\nu}\sum\limits_{\omega}(Z^-_{\omega}Z^{++}_{\omega}E_\pi(\nu,\omega)-Z^+_{\omega}Z^{--}_{\omega}E_\nu(\pi,\omega))],
\label{opn4}	
	\end{eqnarray}
	\begin{eqnarray}
			<{\pi_{corr}}|\tau_+\sigma_{\mu}|{\nu_{corr}}>=q^V_{\pi\nu}+2\chi_{GT}[q^V_{\pi\nu}\sum\limits_{\omega}(Z^{+2}_{\omega}E_\pi(\nu,\omega)+Z^{-2}_{\omega}E_\nu(\pi,\omega)) \nonumber\\
			-q^U_{\pi\nu}\sum\limits_{\omega}Z^-_{\omega}Z^+_{\omega}(E_\pi(\nu,\omega)+E_\nu(\pi,\omega))]
			+2\kappa_{GT}[\tilde{q}_{\pi\nu}\sum\limits_{\omega}(Z^+_{\omega}Z^{++}_{\omega}E_\pi(\nu,\omega)
			-Z^-_{\omega}Z^{--}_{\omega}E_\nu(\pi,\omega))\nonumber\\-q_{\pi\nu}\sum\limits_{\omega}(Z^+_{\omega}Z^{--}_{\omega}E_\pi(\nu,\omega)-Z^-_{\omega}Z^{++}_{\omega}E_\nu(\pi,\omega))],
			\label{opn5}
	\end{eqnarray}
		\begin{equation}
		<{\nu_{corr}}|\tau_{\pm}\sigma_{-\mu}|{\pi_{corr}}>=(-1)^{\mu}<{\pi_{corr}}|\tau_{\mp}\sigma_{\mu}|{\nu_{corr}}>.
		\label{opn6}
	\end{equation}
In Eqs.~(\ref{opn4}), (\ref{opn5}) and (\ref{opn6}), $\sigma_{\mu}$ and $\tau_{\pm}$ are spin and iso-spin type operators,  respectively, and other symbols $q_{\pi\nu}$ ($\tilde{q}_{\pi\nu}$), $q^U_{\pi\nu}$ ($q^V_{\pi\nu}$), $Z^-_{\omega}$ ($Z^+_{\omega}$) and $Z^{--}_{\omega}$ ($Z^{++}_{\omega}$) are defined as
\begin{eqnarray}
q_{\pi\nu}=f_{\pi\nu}u_\pi v_\nu,~~~~ q_{\pi\nu}^{U}=f_{\pi\nu}u_\pi u_\nu, \nonumber \\
\tilde q_{\pi\nu}=f_{\pi\nu}v_\pi u_\nu,~~~~_{\pi\nu}^{V}=f_{\pi\nu}v_\pi v_\nu \nonumber \\
	Z^{-}_{\omega}= \sum_{\pi,\nu}(X^{\pi\nu}_{\omega}q_{\pi\nu}-Y^{\pi\nu}_{\omega}\tilde q_{\pi\nu}),\nonumber \\
	Z^{+}_{\omega}= \sum_{\pi,\nu}(X^{\pi\nu}_{\omega}\tilde q_{\pi\nu}-Y^{\pi\nu}_{\omega}q_{\pi\nu}),
\nonumber \\
	Z^{--}_{\omega}= \sum_{\pi,\nu}(X^{\pi\nu}_{\omega}q^{U}_{\pi\nu}+Y^{\pi\nu}_{\omega}q^{V}_{\pi\nu}),
\nonumber \\
	Z^{+ +}_{\omega}=
	\sum_{\pi\nu}(X^{\pi,\nu}_{\omega}q^{V}_{\pi\nu}+Y^{\pi\nu}_{\omega}q^{U}_{\pi\nu}).
\end{eqnarray}
The terms $X^{\pi\nu}_{\omega}$ and $Y^{\pi\nu}_{\omega}$ were defined earlier and other symbols have usual meanings.
The idea of quasiparticle transitions with first-order phonon correlations can be extended to an odd-odd parent nucleus.
The ground state is assumed to be a proton-neutron quasiparticle pair state of the smallest energy. The GT transitions
of the quasiparticle lead to two-proton or two-neutron quasiparticle states in the even-even daughter nucleus. The
two quasiparticle states were constructed with phonon correlations and given by
	\begin{eqnarray}
			|{\pi \nu_{corr}}>~=~a_\pi^\dagger a^\dagger_\nu|{-}>+\frac{1}{2}\sum\limits_{\pi'_1,\pi'_2,\omega}a^\dagger_{\pi'_1}a^\dagger_{\pi'_2}A^\dagger_{\omega}(-\mu)|{-}>~~~~~~~~~~~~~~~\nonumber\\\times <{-}|[a^\dagger_{\pi'_1}a^\dagger_{\pi'_2}A^\dagger_{\omega}(-\mu)]^\dagger H_{31}a^\dagger_\pi a^\dagger_\nu|{-}>E_{\pi\nu}(\pi'_1\pi'_2,\omega)+\frac{1}{2}\sum\limits_{\nu'_1,\nu'_2,\omega}a^\dagger_{\nu'_1}a_{\nu'_2}A^\dagger_{\omega}(\mu)|{-}>\nonumber\\\times <{-}|[a^\dagger_{\nu'_1}a^\dagger_{\nu'_2}A^\dagger_{\omega}(\mu)]^\dagger
			H_{31}a^\dagger_\pi a^\dagger_\nu|{-}>E_{\pi\nu}(\nu'_1\nu'_2,\omega),
			\label{opn7}
	\end{eqnarray}
	\begin{eqnarray}
			<{\pi_1\pi_{2corr}}|~=~a^\dagger_{\pi_1}a^\dagger_{\pi_2}|{-}>+\sum\limits_{\pi',\nu',\omega}a^\dagger_{\pi'}a^\dagger_{\nu'}A^\dagger_{\omega}(\mu)|{-}>~~~~~~~~~~~~~~~\nonumber\\\times <{-}|[a^\dagger_{\pi'}a^\dagger_{\nu'}A^\dagger_{\omega}(\mu)]^\dagger
			H_{31}a^\dagger_{\pi_1}a^\dagger_{\pi_2}|{-}>E_{\pi_1\pi_2}(\pi'\nu',\omega),
			\label{opn8}
	\end{eqnarray}
	\begin{eqnarray}
			<{\nu_1\nu_{2corr}}|~=~a^\dagger_{\nu_1}a^\dagger_{\nu_2}|{-}>+\sum\limits_{\pi',\nu',\omega}a^+_{\pi'}a^\dagger_{\nu'}A^\dagger_{\omega}(-\mu)|{-}>~~~~~~~~~~~~~~~\nonumber\\\times <{-}|[a^\dagger_{\pi'}a^\dagger_{\nu'}A^\dagger_{\omega}(-\mu)]^\dagger
			H_{31}a^\dagger_{\nu_1}a^\dagger_{\nu_2}|{-}>E_{\nu_1\nu_2}(\pi'\nu',\omega),
			\label{opn9}
	\end{eqnarray}
	where,
		\begin{equation}
			E_{ab}(cd,\omega)=\frac{1}{(\epsilon_a+\epsilon_b)-(\epsilon_{c}+\epsilon_{d}+\omega)}
			\label{opn10}
		\end{equation}
with subscript index a (b) denotes $\pi,~\pi_1$ and $\nu_1$ ($\nu,~\pi_2$ and $\nu_2$) and c (d) denotes
$\pi',~\pi'_1$ and $\nu'_1$ ($\nu',~\pi'_2$ and $\nu'_2$).
The GT transition amplitudes between these states were reduced to those of one quasiparticle states
	\begin{eqnarray}
	<{\pi_1\pi_{2corr}}|\tau_{\pm}\sigma_{\mu}|{\pi \nu_{corr}}>~=~&\delta(\pi_1,\pi)<{\pi_{2corr}}|\tau_{\pm}\sigma_{\mu}|{\nu_{corr}}>\nonumber\\ &-\delta(\pi_2,\pi)
			<{\pi_{1corr}}|\tau_{\pm}\sigma_{\mu}|{\nu_{corr}}>,
			\label{opn11}
		\end{eqnarray}
	\begin{eqnarray}
			<{\nu_1\nu_{2corr}}|\tau_{\pm}\sigma_{-\mu}|{\pi \nu_{corr}}>~=~&\delta(\nu_2,\nu)<{\nu_{1corr}}|\tau_{\pm}\sigma_{-\mu}|{\pi_{corr}}>\nonumber\\ &-\delta(\nu_1,\nu)
			<{\nu_{2corr}}|\tau_{\pm}\sigma_{-\mu}|{\pi_{corr}}>,
			\label{opn12}
	\end{eqnarray}
by ignoring terms of second order in the correlated phonons.
For odd-odd parent nuclei, QRPA phonon excitations are also possible where the quasiparticle pair acts as spectators in the same single quasiparticle shells. The nuclear excited states can be constructed as phonon correlated multi quasiparticle states. The transition amplitudes between the multi quasiparticle states can be reduced to those of one
quasiparticle states as described below.

Excited levels of an even-even nucleus are two-proton quasiparticle states and two-neutron quasiparticle states.
Transitions from these initial states to final neutron-proton quasiparticle pair states are possible
in the odd-odd daughter nuclei. The transition amplitudes can be reduced to correlated quasiparticle states
by taking the Hermitian conjugate of Eq.~(\ref{opn11}) and (\ref{opn12})
\begin{eqnarray}
		<{\pi \nu_{corr}}|\tau_{\pm}\sigma_{-\mu}|{\pi_1\pi_{2corr}}>~=~& - \delta(\pi,\pi_2)<{\nu_{corr}}|\tau_{\pm}\sigma_{-\mu}|{\pi_{1corr}}>\nonumber\\&+\delta(\pi,\pi_1)
		<{\nu_{corr}}|\tau_{\pm}\sigma_{-\mu}|{\pi_{2corr}}>,
		\label{opn13}
\end{eqnarray}
\begin{eqnarray}
		<{\pi \nu_{corr}}|\tau_{\pm}\sigma_{\mu}|{\nu_1\nu_{2corr}}>~=~&\delta(\nu,\nu_2)<{\pi_{corr}}|\tau_{\pm}\sigma_{\mu}|{\nu_{1corr}}>\nonumber\\&-\delta(\nu,\nu_1)
		<{\pi_{corr}}|\tau_{\pm}\sigma_{\mu}|{\nu_{2corr}}>.
		\label{opn13}
\end{eqnarray}

When a nucleus has an odd nucleon (a proton and/or a neutron), low-lying states were obtained by lifting
the quasiparticle in the orbit of the smallest energy to higher-lying orbits. States of an odd-proton even-neutron nucleus
were expressed by three-proton states or one proton two-neutron states, corresponding to excitation
of a proton or a neutron
	\begin{eqnarray}\label{opn14}
		|\pi_1\pi_2\pi_{3corr}\rangle~=~a^\dagger_{\pi_1}a^\dagger_{\pi_2}a^\dagger_{\pi_3}|-\rangle + \frac{1}{2}\sum_{\pi^{'}_1,\pi^{'}_2,\nu^{'},\omega}a^\dagger_{\pi^{'}_1}a^\dagger_{\pi^{'}_2}a^\dagger_{\nu^{'}}A^\dagger_{\omega}(\mu)|-\rangle \nonumber \\
		~~~~~~~~~~~~~~~ \times \langle-|[a^\dagger_{\pi^{'}_1}a^\dagger_{\pi^{'}_2}a^\dagger_{\nu^{'}}A^\dagger_{\omega}(\mu)]^{\dagger}H_{31}a^\dagger_{\pi_1}a^\dagger_{\pi_2}a^\dagger_{\pi_3}|-\rangle \nonumber \\
		~~~~~~~~~~~~~~~ \times E_{\pi_1\pi_2\pi_3}(\pi^{'}_1\pi^{'}_2\nu^{'},\omega)
	\end{eqnarray}
	\begin{eqnarray}\label{opn15}
		|\pi_1\nu_1\nu_{2corr}\rangle ~=~a^\dagger_{\pi_1}a^\dagger_{\nu_1}a^\dagger_{\nu_2}|-\rangle \nonumber + \frac{1}{2}\sum_{\pi^{'}_1,\pi^{'}_2,\nu^{'},\omega}a^\dagger_{\pi^{'}_1}a^\dagger_{\pi^{'}_2}a^\dagger_{\nu^{'}}A^\dagger_{\omega}(-\mu)|-\rangle \nonumber \\
		~~~~~~~~~~~~~~~ \times \langle-|[a^\dagger_{\pi^{'}_1}a^\dagger_{\pi^{'}_2}a^\dagger_{\nu^{'}}A^\dagger_{\omega}(-\mu)]^{\dagger}H_{31}a^\dagger_{\pi_1}a^\dagger_{\nu_1}a^\dagger_{\nu_2}|-\rangle \nonumber \\
		~~~~~~~~~~~~~~~ \times E_{\pi_1\nu_1\nu_2}(\pi^{'}_1\pi^{'}_2\nu^{'},\omega) \nonumber +\frac{1}{6}\sum_{\nu^{'}_1,\nu^{'}_2,\nu^{'}_3,\omega}a^\dagger_{\nu^{'}_1}a^\dagger_{\nu^{'}_2}a^\dagger_{\nu^{'}_3}A^\dagger_{\omega}(\mu)|-\rangle \nonumber \\
		~~~~~~~~~~~~~~~ \times \langle-|[a^\dagger_{\nu^{'}_1}a^\dagger_{\nu^{'}_2}a^\dagger_{\nu^{'}_3}A^\dagger_{\omega}(\mu)]^{\dagger}H_{31}a^\dagger_{\pi_1}a^\dagger_{\nu_1}a^\dagger_{\nu_2}|-\rangle \nonumber \\
		~~~~~~~~~~~~~~~ \times E_{\pi_1\nu_1\nu_2}(\nu^{'}_1\nu^{'}_2\nu^{'}_3,\omega)
	\end{eqnarray}
	with the energy denominators of first order perturbation,
	\begin{equation}
			E_{abc}(def,\omega)=\frac{1}{(\epsilon_{a}+\epsilon_{b}+\epsilon_{c}-\epsilon_{d}-\epsilon_{e}-\epsilon_{f}-\omega)},
			\label{opn16}
		\end{equation}
where subscripts represent $\pi_1$, $\pi_2$, $\pi_3$, $\nu_1$ and $\nu_2$ ($\pi'_1$, $\pi'_2$, $\nu'$, $\nu'_1$, $\nu'_2$ and $\nu'_2$). These equations can be used to generate the three quasiparticle states of odd-proton and even-neutron by swapping the neutron and proton states, $\nu\longleftrightarrow \pi$ and $A^{\dagger}_\omega(\mu) \longleftrightarrow A^{\dagger}_\omega(-\mu)$. Amplitudes of the quasiparticle transitions between the three quasiparticle states were reduced to those for correlated one quasiparticle states. For parent nuclei with an odd proton,
\begin{eqnarray}
		\langle \pi^{'}_1\pi^{'}_2\nu^{'}_{1corr}|\tau_{\pm}\sigma_{-\mu}|\pi_1\pi_2\pi_{3corr}\rangle&
		~=~\delta(\pi^{'}_1,\pi_2)\delta(\pi^{'}_2,\pi_3)\langle \nu^{'}_{1corr}|\tau_{\pm}\sigma_{-\mu}|\pi_{1corr}\rangle \nonumber\\
		~&-\delta(\pi^{'}_1,\pi_1)\delta(\pi^{'}_2,\pi_3)\langle \nu^{'}_{1corr}|\tau_{\pm}\sigma_{-\mu}|\pi_{2corr}\rangle \nonumber\\
		~&+\delta(\pi^{'}_1,\pi_1)\delta(\pi^{'}_2,\pi_2)\langle \nu^{'}_{1corr}|\tau_{\pm}\sigma_{-\mu}|\pi_{3corr}\rangle,
	\end{eqnarray}
	\begin{eqnarray}
		\langle \pi^{'}_1\pi^{'}_2\nu^{'}_{1corr}|\tau_{\pm}\sigma_{\mu}|\pi_1\nu_1\nu_{2corr}\rangle&
		~=~\delta(\nu^{'}_1,\nu_2)[\delta(\pi^{'}_1,\pi_1)\langle \pi^{'}_{2corr}|\tau_{\pm}\sigma_{\mu}|\nu_{1corr}\rangle \nonumber\\
		~&-\delta(\pi^{'}_2,\pi_1)\langle \pi^{'}_{1corr}|\tau_{\pm}\sigma_{\mu}|\nu_{1corr}\rangle] \nonumber\\
		~&-\delta(\nu^{'}_1,\nu_1)[\delta(\pi^{'}_1,\pi_1)\langle \pi^{'}_{2corr}|\tau_{\pm}\sigma_{\mu}|\nu_{2corr}\rangle \nonumber\\
		~&-\delta(\pi^{'}_2,\pi_1)\langle \pi^{'}_{1corr}|\tau_{\pm}\sigma_{\mu}|\nu_{2corr}\rangle],
	\end{eqnarray}
	\begin{eqnarray}
		\langle \nu^{'}_1\nu^{'}_2\nu^{'}_{3corr}|\tau_{\pm}\sigma_{-\mu}|\pi_1\nu_1\nu_{2corr}\rangle &
		~=~\delta(\nu^{'}_2,\nu_1)\delta(\nu^{'}_3,\nu_2)\langle \nu^{'}_{1corr}|\tau_{\pm}\sigma_{-\mu}|\pi_{1corr}\rangle \nonumber\\
		~&-\delta(\nu^{'}_1,\nu_1)\delta(\nu^{'}_3,\nu_2)\langle \nu^{'}_{2corr}|\tau_{\pm}\sigma_{-\mu}|\pi_{1corr}\rangle \nonumber\\
		~&+\delta(\nu^{'}_1,\nu_1)\delta(\nu^{'}_2,\nu_2)\langle \nu^{'}_{3corr}|\tau_{\pm}\sigma_{-\mu}|\pi_{1corr}\rangle,
	\end{eqnarray}
	and for parent nuclei with an odd neutron
	\begin{eqnarray}
		\langle \pi^{'}_1\nu^{'}_1\nu^{'}_{2corr}|\tau_{\pm}\sigma_{\mu}|\nu_1\nu_2\nu_{3corr}\rangle
		&~=~\delta(\nu^{'}_1,\nu_2)\delta(\nu^{'}_2,\nu_3)\langle \pi^{'}_{1corr}|\tau_{\pm}\sigma_{\mu}|\nu_{1corr}\rangle \nonumber\\
		~&-\delta(\nu^{'}_1,\nu_1)\delta(\nu^{'}_2,\nu_3)\langle \pi^{'}_{1corr}|\tau_{\pm}\sigma_{\mu}|\nu_{2corr}\rangle \nonumber\\
		~&+\delta(\nu^{'}_1,\nu_1)\delta(\nu^{'}_2,\nu_2)\langle \pi^{'}_{1corr}|\tau_{\pm}\sigma_{\mu}|\nu_{3corr}\rangle,
	\end{eqnarray}
	\begin{eqnarray}
		\langle \pi^{'}_1\nu^{'}_1\nu^{'}_{2corr}|\tau_{\pm}\sigma_{-\mu}|\pi_1\pi_2\nu_{1corr}\rangle
		&~=~\delta(\pi^{'}_1,\pi_2)[\delta(\nu^{'}_1,\nu_1)\langle \nu^{'}_{2corr}|\tau_{\pm}\sigma_{-\mu}|\pi_{1corr}\rangle \nonumber\\
		&~-\delta(\nu^{'}_2,\nu_1)\langle \nu^{'}_{1corr}|\tau_{\pm}\sigma_{-\mu}|\pi_{1corr}\rangle] \nonumber\\
		~&-\delta(\pi^{'}_1,\pi_1)[\delta(\nu^{'}_1,\nu_1)\langle \nu^{'}_{2corr}|\tau_{\pm}\sigma_{-\mu}|\pi_{2corr}\rangle \nonumber\\
		~&-\delta(\nu^{'}_2,\nu_1)\langle \nu^{'}_{1corr}|\tau_{\pm}\sigma_{-\mu}|\pi_{2corr}\rangle],
	\end{eqnarray}
	\begin{eqnarray}
		\langle \pi^{'}_1\pi^{'}_2\pi^{'}_{3corr}|\tau_{\pm}\sigma_{\mu}|\pi_1\pi_2\nu_{1corr}\rangle
		&~=~\delta(\pi^{'}_2,\pi_1)\delta(\pi^{'}_3,\pi_2)\langle \pi^{'}_{1corr}|\tau_{\pm}\sigma_{\mu}|\nu_{1corr}\rangle \nonumber\\
		&~-\delta(\pi^{'}_1,\pi_1)\delta(\pi^{'}_3,\pi_2)\langle \pi^{'}_{2corr}|\tau_{\pm}\sigma_{\mu}|\nu_{1corr}\rangle \nonumber\\
		~&+\delta(\pi^{'}_1,\pi_1)\delta(\pi^{'}_2,\pi_2)\langle \pi^{'}_{3corr}|\tau_{\pm}\sigma_{\mu}|\nu_{1corr}\rangle.
	\end{eqnarray}

Low-lying states in an odd-odd nucleus were expressed
in the quasiparticle picture by proton-neutron pair states (two quasiparticle states) or by states which were
obtained by adding two proton or two-neutron quasi-particles (four quasiparticle states). Transitions from the former
states were described earlier. Phonon-correlated four quasiparticle states can be constructed
similarly to the two and three quasiparticle states. Also in this
case, transition amplitudes for the four quasiparticle states were
reduced into those for the correlated one quasiparticle states	
	\begin{eqnarray}
			<{\pi^{'}_1\pi^{'}_2\nu^{'}_1\nu^{'}_{2corr}}|\tau_{\pm}\sigma_{-\mu}|{\pi_1\pi_2\pi_3\nu_{1corr}}
>~~~~~~~~~~~~~~~~~~~~~~~~~~~~~~~~~~~~~~~~~~~~~~~~~~~~~~~~~~~~~~~~~~~~~~~\nonumber\\~~~~~~~~~~~~~~=~\delta(\nu^{'}_2,\nu_1)[\delta(\pi^{'}_1,\pi_2)\delta(\pi^{'}_2,\pi_3)<{\nu^{'}_{1corr}}|\tau_{\pm}\sigma_{-\mu}|{\pi_{1corr}}>\nonumber\\
			-\delta(\pi^{'}_1,\pi_1)\delta(\pi^{'}_2,\pi_3)<{\nu^{'}_{1corr}}|\tau_{\pm}\sigma_{-\mu}|{\pi_{2corr}}>\nonumber\\+\delta(\pi^{'}_1,\pi_1)\delta(\pi^{'}_2,\pi_2)<{\nu^{'}_{1corr}}|\tau_{\pm}\sigma_{-\mu}|{\pi_{3corr}}>]\nonumber\\
			-\delta(\nu^{'}_1,\nu_1)[\delta(\pi^{'}_1,\pi_2)\delta(\pi^{'}_2,\pi_3)<{\nu^{'}_{2corr}}|\tau_{\pm}\sigma_{-\mu}|{\pi_{1corr}}>\nonumber\\
			-\delta(\pi^{'}_1,\pi_1)\delta(\pi^{'}_2,\pi_3)<{\nu^{'}_{2corr}}|\tau_{\pm}\sigma_{-\mu}|{\pi_{2corr}}>\nonumber\\
			+\delta(\pi^{'}_1,\pi_1)\delta(\pi^{'}_2,\pi_2)<{\nu^{'}_{2corr}}|\tau_{\pm}\sigma_{-\mu}|{\pi_{3corr}}>],
		\end{eqnarray}
	\begin{eqnarray}
			<{\pi^{'}_1\pi^{'}_2\pi^{'}_3\pi^{'}_{4corr}}|\tau_{\pm}\sigma_{\mu}|{\pi_1\pi_2\pi_3\nu_{1corr}}>~~~~~~~~~~~~~~~~~~~~~~~~~~~~~~~~~~~~~~~~~~~~~~~~~~~~~~~~~~~~~~~~~~~~~~~\nonumber\\~~~~~~~~~~~~~~~=~~-\delta(\pi^{'}_2,\pi_1)\delta(\pi^{'}_3,\pi_2)\delta(\pi^{'}_4,\pi_3)<{\pi^{'}_{1corr}}|\tau_{\pm}\sigma_{\mu}|{\nu_{1corr}}>\nonumber\\
			+\delta(\pi^{'}_1,\pi_1)\delta(\pi^{'}_3,\pi_2)\delta(\pi^{'}_4,\pi_3)<{\pi^{'}_{2corr}}|\tau_{\pm}\sigma_{\mu}|{\nu_{1corr}}>\nonumber\\
			-\delta(\pi^{'}_1,\pi_1)\delta(\pi^{'}_2,\pi_2)\delta(\pi^{'}_4,\pi_3)<{\pi^{'}_{3corr}}|\tau_{\pm}\sigma_{\mu}|{\nu_{1corr}}>\nonumber\\
			+\delta(\pi^{'}_1,\pi_1)\delta(\pi^{'}_2,\pi_2)\delta(\pi^{'}_3,\pi_3)<{\pi^{'}_{4corr}}|\tau_{\pm}\sigma_{\mu}|{\nu_{1corr}}>,
				\end{eqnarray}
	\begin{eqnarray}
			<{\pi^{'}_1\pi^{'}_2\nu^{'}_1\nu^{'}_{2corr}}|\tau_{\pm}\sigma_{\mu}|{\pi_1\nu_1\nu_2\nu_{3corr}}>~~~~~~~~~~~~~~~~~~~~~~~~~~~~~~~~~~~~~~~~~~~~~~~~~~~~~~~~~~~~~~~~~~~~~~~\nonumber\\~~~~~~~~~~~~~~~=~~\delta(\pi^{'}_1,\pi_1)[\delta(\nu^{'}_1,\nu_2)\delta(\nu^{'}_2,\nu_3)<{\pi^{'}_{2corr}}|\tau_{\pm}\sigma_{\mu}|{\nu_{1corr}}>\nonumber\\
			-\delta(\nu^{'}_1,\nu_1)\delta(\nu^{'}_2,\nu_3)<{\pi^{'}_{2corr}}|\tau_{\pm}\sigma_{\mu}|{\nu_{2corr}}>\nonumber\\
			+\delta(\nu^{'}_1,\nu_1)\delta(\nu^{'}_2,\nu_2)<{\pi^{'}_{2corr}}|\tau_{\pm}\sigma_{\mu}|{\nu_{3corr}}>]\nonumber\\
			-\delta(\pi^{'}_2,\pi_1)[\delta(\nu^{'}_1,\nu_2)\delta(\nu^{'}_2,\nu_3)<{\pi^{'}_{1corr}}|\tau_{\pm}\sigma_{\mu}|{\nu_{1corr}}>\nonumber\\
			-\delta(\nu^{'}_1,\nu_1)\delta(\nu^{'}_2,\nu_3)<{\pi^{'}_{1corr}}|\tau_{\pm}\sigma_{\mu}|{\nu_{2corr}}>\nonumber\\
			+\delta(\nu^{'}_1,\nu_1)\delta(\nu^{'}_2,\nu_2)<{\pi^{'}_{1corr}}|\tau_{\pm}\sigma_{\mu}|{\nu_{3corr}}>],
	\end{eqnarray}
	\begin{eqnarray}
			<{\nu^{'}_1\nu^{'}_2\nu^{'}_3\nu^{'}_{4corr}}|\tau_{\pm}\sigma_{-\mu}|{\pi_1\nu_1\nu_2\nu_{3corr}}>~~~~~~~~~~~~~~~~~~~~~~~~~~~~~~~~~~~~~~~~~~~~~~~~~~~~~~~~~~~~~~~~~~~~~~~\nonumber\\~~~~~~~~~~~~~~~=~~+\delta(\nu^{'}_2,\nu_1)\delta(\nu^{'}_3,\nu_2)\delta(\nu^{'}_4,\nu_3)<{\nu^{'}_{1corr}}|\tau_{\pm}\sigma_{-\mu}|{\pi_{1corr}}>\nonumber\\
			-\delta(\nu^{'}_1,\nu_1)\delta(\nu^{'}_3,\nu_2)\delta(\nu^{'}_4,\nu_3)<{\nu^{'}_{2corr}}|\tau_{\pm}\sigma_{-\mu}|{\pi_{1corr}}>\nonumber\\
			+\delta(\nu^{'}_1,\nu_1)\delta(\nu^{'}_2,\nu_2)\delta(\nu^{'}_4,\nu_3)<{\nu^{'}_{3corr}}|\tau_{\pm}\sigma_{-\mu}|{\pi_{1corr}}>\nonumber\\
			-\delta(\nu^{'}_1,\nu_1)\delta(\nu^{'}_2,\nu_2)\delta(\nu^{'}_3,\nu_3)<{\nu^{'}_{4corr}}|\tau_{\pm}\sigma_{-\mu}|{\pi_{1corr}}>.
	\end{eqnarray}
The antisymmetrization of the quasi-particles was duly taken into account for each of these amplitudes.\\
	$\pi^{'}_4>\pi^{'}_3>\pi^{'}_2>\pi^{'}_1$,  $\nu^{'}_4>\nu^{'}_3>\nu^{'}_2>\nu^{'}_1$,  $\pi_4>\pi_3>\pi_2>\pi_1$,  $\nu_4>\nu_3>\nu_2>\nu_1$.\\
	The GT transitions were taken into account for each phonon's excited state. It was assumed that the quasiparticle in the parent nucleus occupies the same orbit as the excited phonons.

The form of the Hamiltonian for many-particle QRPA system is
\begin{equation}
H_{QRPA} = H_{sp} + \hat{V}_{pairing} + \hat{V}_{pp(GT)} + \hat{V}_{ph(GT)},
\label{Ham}
\end{equation}
where $H_{sp}$ is the single-particle Hamiltonian whose energies and wave-vectors were calculated using the deformed Nilsson model~\cite{Nil55}.  $\hat{V}_{pp(GT)}$ (Eq.~\ref{ppGT}) and $\hat{V}_{ph(GT)}$ (Eq.~\ref{phGT}) were introduced earlier in this section. The pairing correlations ($\hat{V}_{pairing}$) were taken into account within the BCS formalism with fixed pairing gaps between proton-proton ($\Delta_{\pi\pi}$) and neutron-neutron ($\Delta_{\nu\nu}$) systems. The values of pairing gaps were calculated  using empirical formulae~\cite{Wan12} between neutron-neutron ($\Delta_{\nu\nu}$) and proton-proton ($\Delta_{\pi\pi}$) systems. The expressions for these gaps were given by 
\begin{eqnarray}
   \Delta_{\nu\nu} &=& \frac{(-1)^{1-Z+A}[S_{\nu}(A-1,Z)-2S_{\nu}(A,Z)+S_{\nu}(A+1, Z)]}{4}, \nonumber \\
    \Delta_{\pi\pi} &=& \frac{(-1)^{1+Z}[S_{\pi}(A-1,Z-1)-2S_{\pi}(A,Z)+S_{\pi}(A+1, Z+1)]}{4},
\end{eqnarray}
where the proton and neutron separation energies,  $S_{\pi}$ and  $S_{\nu}$, respectively, were taken from~\cite{Mol19} for cases where latest experimental data~\cite{nds} were not available.
Nuclear deformation values were taken from~\cite{Mol16}, mass excess values were adopted from~\cite{nds},  Nilsson oscillatory constant was chosen as
\begin{equation}
	\Omega=\frac{45}{A^{1/3}}-\frac{25}{A^{2/3}},
	\label{Eq:NOC}
\end{equation}
with same values for neutrons and protons, and Nilsson potential parameters were used to compute the weak rates. 

We calculated both U1F and allowed transitions in this work. The allowed transitions depend only on spin ($\sigma_{\mu}$) and iso-spin ($\tau_{\pm}$) type operators, while forbidden transitions also contain $rY_{lm}$ where $Y_{lm}$ are the associated spherical harmonics.

The matrix elements of U1F transitions in pp and ph directions were given by
\begin{equation}
V^{pp}_{\pi\nu,\pi^{\prime}\nu^{\prime}} = -2\kappa_{U1F}f_{\pi\nu}(\mu)f_{\pi^{\prime}\nu^{\prime}}(\mu),
\label{ppF}
\end{equation}
\begin{equation}
V^{ph}_{\pi\nu,\pi^{\prime}\nu^{\prime}} = 2\chi_{U1F}f_{\pi\nu}(\mu)f_{\pi^{\prime}\nu^{\prime}}(\mu),
\label{phF}
\end{equation}
where ph and pp interaction constants are, respectively referred to as $\chi_{U1F}$ and $\kappa_{U1F}$ and single-particle amplitude ($f_{\pi\nu}(\mu)$) of U1F transition is given by
\begin{equation}
f_{\pi\nu}(\mu)=\langle \pi|\tau_{-}r[\sigma Y_{1}]_{2\mu}|\nu \rangle,
\end{equation}
where parities of neutron ($|\nu\rangle$)
and proton ($|\pi\rangle$) states are opposite to each other~\cite{Hom96} and $\mu$ takes values $0,\pm1$, and $\pm2$. The 
parametrization of ph and pp strength interaction constants, for both allowed and U1F transitions, 
were adopted from~\cite{Hom96}.

The partial decay rate ($\lambda_{if}$) for any transition between the parent ($i$) and the daughter ($f$) states was calculated using
\begin{eqnarray}
\lambda_{if} = \left(\frac{m^{5}_{e}c^{4}g^{2}}{2\hbar^{7} \pi^{3}} \right) \Phi_{if}(E_{fermi}, T, \rho) B_{if},
\label{Eq:Prate}
\end{eqnarray} 
having dependence upon the $g$ (weak coupling constant) involving both vector ($g_{V}$) and axial-vector ($g_{A}$) type constants, $B_{if}$ (reduced transition probabilities) and $\Phi_{if}$ (phase-space integrals). 
In the case of continuum (allowed) EC these integrals was computed using
\begin{eqnarray}
\Phi^{EC}_{ij} = \int _{w_{l}}^{\infty}w(w^{2} -1)^{1/2}(w_{m}+w)^{2}(G_{-})F(+Z,w)\;dw,
\label{Eq:FIPC}
\end{eqnarray}
while for allowed BD was calculated as follows
\begin{eqnarray}
\Phi^{BD}_{ij} = \int_{1}^{w_{m}}w({w^{2} -1)^{1/2}(w_{m}-w)^{2}(1-G_{-})F(+Z,w)}\;dw,
\label{Eq:FIEE}
\end{eqnarray}

In the case of U1F transition, the expression of phase-space integrals ($\Phi^{U1F}_{if}$) is given below
\begin{eqnarray}
\Phi^{U1F}_{ij} =& \int _{1}^{w_{m}}\{w
(w^{2} -1)^{1/2} (w_{m}-w)^{2} (1-G_{-}) [F_{1}(Z,w) \nonumber \\
&(w_{m}-w)^{2}+F_{2}(Z,w)(w^{2}-1)]\}\;dw.
\label{Eq:UFI}
\end{eqnarray}
In Eqs.~(\ref{Eq:FIPC}~-~\ref{Eq:UFI}), we used natural units ($\hbar=m_{e}=c=1$). $w$ denotes the total electron energy, which includes the kinetic and rest mass energies, and $w_{l}$ signifies the energy threshold for EC, while  $w_{m}$ represents the total BD energy. The symbol $G_{-}$ denotes the Fermi 
Dirac distribution function for electrons. The Fermi functions (F, F$_{1}$ and F$_{2}$) used in this study were adopted from~\cite{Gove71}.  

Reduced transition probability ($B_{if}$) is
\begin{eqnarray}
B_{if} &=& \left(\frac{g_{A}}{g_{V}}\right)^{2}B(GT_{\pm})_{if} + B(F_{\pm})_{if},
\label{Eq:Rprob}
\end{eqnarray}
where $B(GT_{\pm})_{if}$ and $B(F_{\pm})_{if}$ are GT and Fermi transition probabilities, respectively. In Eq. (\ref{Eq:Rprob}) 
value of $\frac{g_{A}}{g_{V}} = -1.2694$ (taken from~\cite{Nak10}). The expressions for these probabilities are
\begin{eqnarray}
B(F_{\pm})_{if} &=& \frac{|\langle f||\hat{O}||i \rangle|^{2}}{2J_{i}+1}; \qquad \hat{O} = \sum_{l}\tau^{l}_{\pm},
\label{Eq:GTP}
\end{eqnarray}
and
\begin{eqnarray}
B(GT_{\pm})_{if} &=& \frac{|\langle f||\hat{O}||i \rangle|^{2}}{2J_{i}+1}; \qquad \hat{O} = \sum_{l}\tau^{l}_{\pm}\bsigma^{l},
\label{Eq:FTP}
\end{eqnarray}
where symbols have usual meanings. For U1F transition, the reduced probability is
\begin{eqnarray}
B(U1F)_{if} =\frac{1}{6}\eta^{2}w^{2}-\frac{1}{6}\eta^{2}w_{m}w+\frac{1}{12}\eta^{2}(w^{2}_{m}-1),
\label{Eq:UFTP}
\end{eqnarray}
where
\begin{eqnarray}
\eta = 2g_{A}(2J_{i}+1)^{-1/2}\langle f||\sum_{l}r_{l}[C^{l}_{1}\times\vec{\sigma}]^{2}\tau^{l}_{-}||i \rangle, \nonumber \\ C_{kk^{'}} = \left(\frac{4\pi}{2l+1}\right)^{1/2}\mathbf{Y}_{kk^{'}},
\label{Eq:z}
\end{eqnarray}

The partial rates were summed over all states in parent as well as daughter nuclei to acquire required convergence in the rates. 
The expression for total rate is given by 
\begin{eqnarray}
\lambda = \sum _{if}P_{i} \lambda_{if},
\label{Eq:Trate}
\end{eqnarray} 
where the excited state occupation probability (P$_{i}$) of parent nuclei were determined by applying the normal Boltzmann distribution. 

In Eq.~(\ref{Eq:Trate}), total rates ($\lambda$) have been determined  using the pn-QRPA formalism, where state-by-state transitions between excited and ground states of parent and daughter nuclei were considered for the calculation of the strength functions in a totally microscopic fashion. The rates based on the \textit{BA} hypothesis were estimated by replicating the strength functions 
for all  parent excited states with ground level strengths~\cite{Fuller}.
Hereafter, microscopic (\textit{Full}) pn-QRPA rates and those based on the \textit{BA} hypothesis would be referred to as 
$\lambda_{F}$  and $\lambda_{BA}$, respectively. 

In order to compare $\lambda_{F}$  and $\lambda_{BA}$ rates, we introduce two new parameters. These are ratios and average deviations of the calculated rates.  The algebraic expressions for the ratio (R$_{i}$) and the average deviation ($\bar{R}$) are
\begin{equation}
	R_i = \left\{ \begin{array}{ll} \lambda_{F}/\lambda_{BA} & \mbox{if $\lambda_{F}\geq \lambda_{BA}$} \\ \\
		
		\lambda_{BA}/\lambda_{F} & \mbox{if $\lambda_{F} < \lambda_{BA}$},
	\end{array}
	\label{Ri}
	\right.
\end{equation}
\begin{equation}
	\bar{R} = \frac{\sum_{i=1}^{k}R_i}{k} ,
	\label{Rbar}
\end{equation}
where $k$ denotes the total count of temperature-density points considered  in the analysis. 

\section{Results and Discussions} \label{sec:results}
\begin{figure}[H]
\begin{center}
\includegraphics[width=0.9\textwidth]{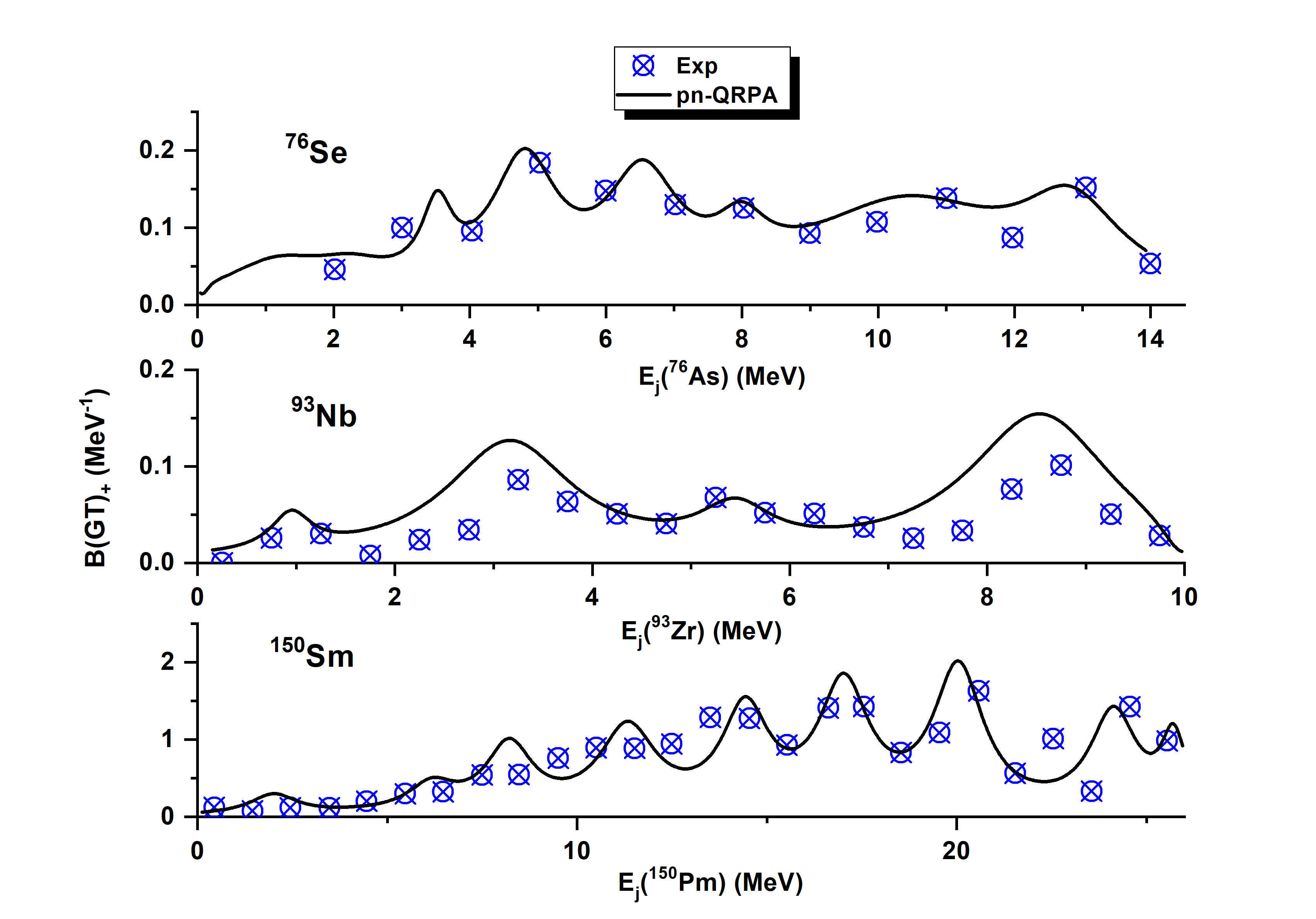}
\end{center}
\caption{\small Comparison of pn-QRPA calculated 
GT$_{+}$ strength distributions of $^{76}$Se, $^{93}$Nb and $^{150}$Sm with measured data taken 
from~\cite{Helmer97},~\cite{Gao20} and~\cite{Gue11}, respectively. 
}\label{Fig. 1}
\end{figure}

\begin{figure}[H]
\begin{center}
\includegraphics[width=0.9\textwidth]{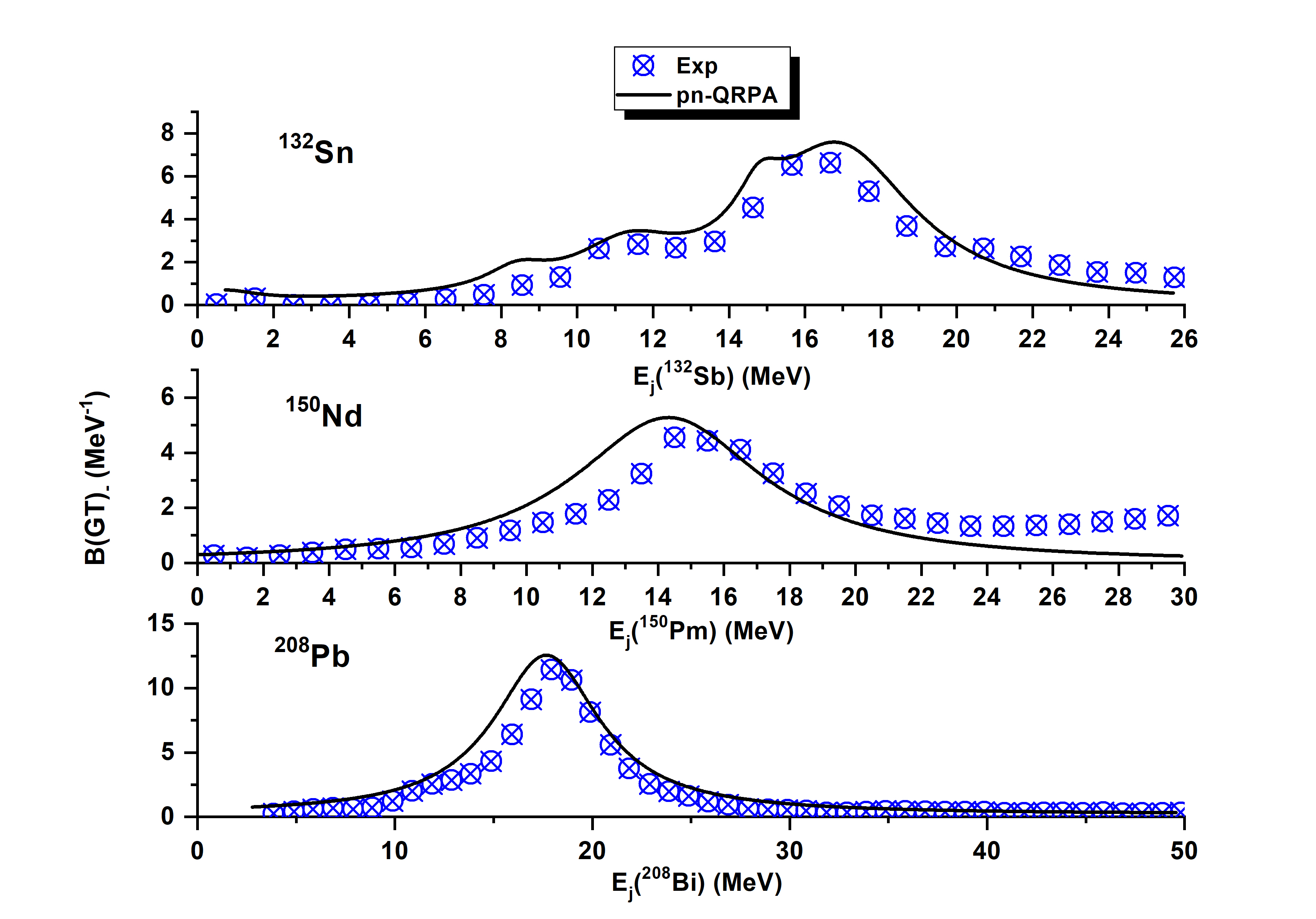}
\end{center}
\caption{\small Comparison of  pn-QRPA calculated 
GT$_{-}$ strength distributions of $^{132}$Sn, $^{150}$Nd and $^{208}$Pb with measured data taken from~\cite{Yas18},~\cite{Gue11} and~\cite{Wak12}, respectively.
}\label{Fig. 2}
\end{figure}

As stated earlier, the aim of this study is to present a quantitative analysis of the reliability of  \textit{BA} rates, specially for U1F transitions. Our work is a continuation of our previous research which focused only on allowed transitions~\cite{Fak23}. We selected a specific region comprising 106 nuclei with A and Z ranging from (70 -- 208) and (27 -- 82), respectively, for the current study. This region is particularly sensitive to the $r$-process. The selected nuclei have been reported 
in theoretical~\cite{Mol03,Mar16,Bor0811,Lan03,Cue07,Zhi13,Bor0305,Zho17,Niu19,Nab14,Nab15,Nabi19,Nabi21,Rob22} and experimental~\cite{Nis11,Lor15,Dom19,Hal21,Pfe02} works. To check reliability of the current model, we first present a comparison 
between the pn-QRPA calculated strength distributions  for allowed GT transitions and measured  data. 
For this purpose, we  applied a smearing technique involving Lorentzian fitting to the theoretical strength distributions with an artificial width based on the calculated spectrum. This technique has been commonly used~\cite{Yas18,Gue11,Wak12,Gao20} to compare the experimental (measured in MeV$^{-1}$ units) and theoretical strength distributions. 
A decent comparison between theory and experiment  can be seen 
from Figs.~(\ref{Fig. 1} - \ref{Fig. 2}) in (GT)$_+$ and (GT)$_-$ directions, respectively. In both of these figures GT strengths [MeV$^{-1}$] are plotted as a function of 
excitation energies [MeV] of the corresponding daughter nuclei (along the abscissa). It is noted that the strength is well fragmented. After establishing the reliability of current model, we next proceed to further our investigation using the pn-QRPA model.

The current and Homma \textit{et al.}~\cite{Hom96} calculations used the same nuclear Hamiltonian in the framework of pn-QRPA with a schematic GT residual interaction. Additionally, incorporation of U1F transitions in our model was done as per recipe given in~\cite{Hom96}. The reliability of the current model for calculation of U1F transitions was discussed earlier in~\cite{Hom96}. Table~\ref{table:table1} reproduces the data shown in~\cite{Hom96} with the latest measured half-lives~\cite{nds}.  A decent agreement between the calculated and measured half-lives is obtained for $^{39}$Ar and $^{41}$Ca, for which $\beta$-decay is known experimentally to be dominated by U1F transitions. It may be seen from  Table~\ref{table:table1} that, for certain cases, taking only the allowed and U1F transitions into account overestimates $\beta$-decay half-lives. This suggests that rank 0 and rank 1 FF transitions possess significant contributions for these nuclei. For the inclusion of non-unique FF transitions in the stellar rate calculations, work on code is currently in progress and we plan to report our findings in the near future.

\begin{table}[h!]
\caption{\small Contribution of U1F transition to total $\beta$-decay for selected nuclei.  
T$_{1/2}^{All}$ is the partial half-life for allowed $\beta$-decay. T$_{1/2}^{total}$ is the total half-life including both allowed and U1F decay. Measured half-lives were taken from \cite{nds}. The dash indicates that the calculation predicts no allowed transition. The table was adopted from \cite{Hom96}.}\label{table:table1} \setlength{\tabcolsep}{20pt}
\centering {
{\small \resizebox{!}{3.5cm}{%
\begin{tabular}{cccccc}

\multicolumn{1}{c}{\bf{Nuclei}} &
\multicolumn{1}{c}{\bf{Decay mode}} &
\multicolumn{1}{c}{\bf{T$_{1/2}^{expt}$ (s)}} &
\multicolumn{1}{c}{\bf{T$_{1/2}^{All}$ (s)}} &
\multicolumn{1}{c}{\bf{T$_{1/2}^{total}$ (s)}} &
\multicolumn{1}{c}{\bf{Contribution (\%)}}  \\
\midrule
    $^{36}$P    & $\beta^{-}$ & 5.60$\times$10$^{+00}$ & 1.38$\times$10$^{+02}$ & 1.36$\times$10$^{+02}$ & 1.7 \\
    $^{37}$S    & $\beta^{-}$ & 3.03$\times$10$^{+02}$ & 1.48$\times$10$^{+02}$ & 1.46$\times$10$^{+02}$ & 1.6 \\
    $^{38}$Cl    & $\beta^{-}$ & 2.23$\times$10$^{+03}$ &    --   & 4.36$\times$10$^{+08}$ & 100.0 \\
    $^{39}$Ar    & $\beta^{-}$ & 8.46$\times$10$^{+09}$ &    --   & 7.60$\times$10$^{+09}$ & 100.0 \\
    $^{41}$Ca    & $\beta^{+}$ & 3.14$\times$10$^{+12}$ &   --    & 7.40$\times$10$^{+11}$ & 100.0 \\
    $^{133}$Sn    & $\beta^{-}$ & 1.46$\times$10$^{+00}$ & 5.07$\times$10$^{+01}$ & 4.50$\times$10$^{+01}$ & 11.4 \\
    $^{134}$Sb    & $\beta^{-}$ & 7.80$\times$10$^{-01}$ & 3.71$\times$10$^{+02}$ & 3.40$\times$10$^{+02}$ & 8.1 \\
    $^{135}$Te    & $\beta^{-}$ & 1.90$\times$10$^{+01}$ & 2.89$\times$10$^{+03}$ & 1.26$\times$10$^{+03}$ & 56.3 \\
    $^{136}$I    & $\beta^{-}$ & 8.34$\times$10$^{+01}$ & 9.57$\times$10$^{+03}$ & 4.96$\times$10$^{+03}$ & 48.2 \\
    $^{137}$Xe    & $\beta^{-}$ & 2.29$\times$10$^{+02}$ & 4.51$\times$10$^{+03}$ & 3.72$\times$10$^{+03}$ & 17.4 \\
    $^{138}$Cs    & $\beta^{-}$ & 1.95$\times$10$^{+03}$ & 6.83$\times$10$^{+04}$ & 3.74$\times$10$^{+04}$ & 45.2 \\
    $^{139}$Ba    & $\beta^{-}$ & 4.98$\times$10$^{+03}$ & 3.74$\times$10$^{+04}$ & 3.55$\times$10$^{+04}$ & 5.0 \\
    $^{140}$La    & $\beta^{-}$ & 1.45$\times$10$^{+05}$ & 8.79$\times$10$^{+04}$ & 8.59$\times$10$^{+04}$ & 2.3 \\
    $^{141}$Ce    & $\beta^{-}$ & 2.81$\times$10$^{+06}$ & --     & 1.31$\times$10$^{+09}$ & 100.0 \\
    $^{142}$Pr    & $\beta^{-}$ & 6.88$\times$10$^{+04}$ &  --     & 9.53$\times$10$^{+11}$ & 100.0 \\
    $^{144}$Pm    & $\beta^{+}$ & 3.14$\times$10$^{+07}$ &   --    & 7.56$\times$10$^{+10}$ & 100.0 \\
 \bottomrule
\end{tabular}}}
}\end{table}

As mentioned earlier, for the nuclei under current investigation, previous works focused only on the calculations and measurements of half-lives and beta-delayed 
neutron-emission probabilities. However, it is of utmost importance to evaluate the reliability of the \textit{BA} hypothesis in calculation of stellar rates for the selected pool of nuclei. With this consideration, two sets of calculation (one for allowed and one for U1F rates) were performed,  separately, for EC and BD decays. 

\begin{figure}[H]
\begin{center}
\includegraphics[width=1.1\textwidth]{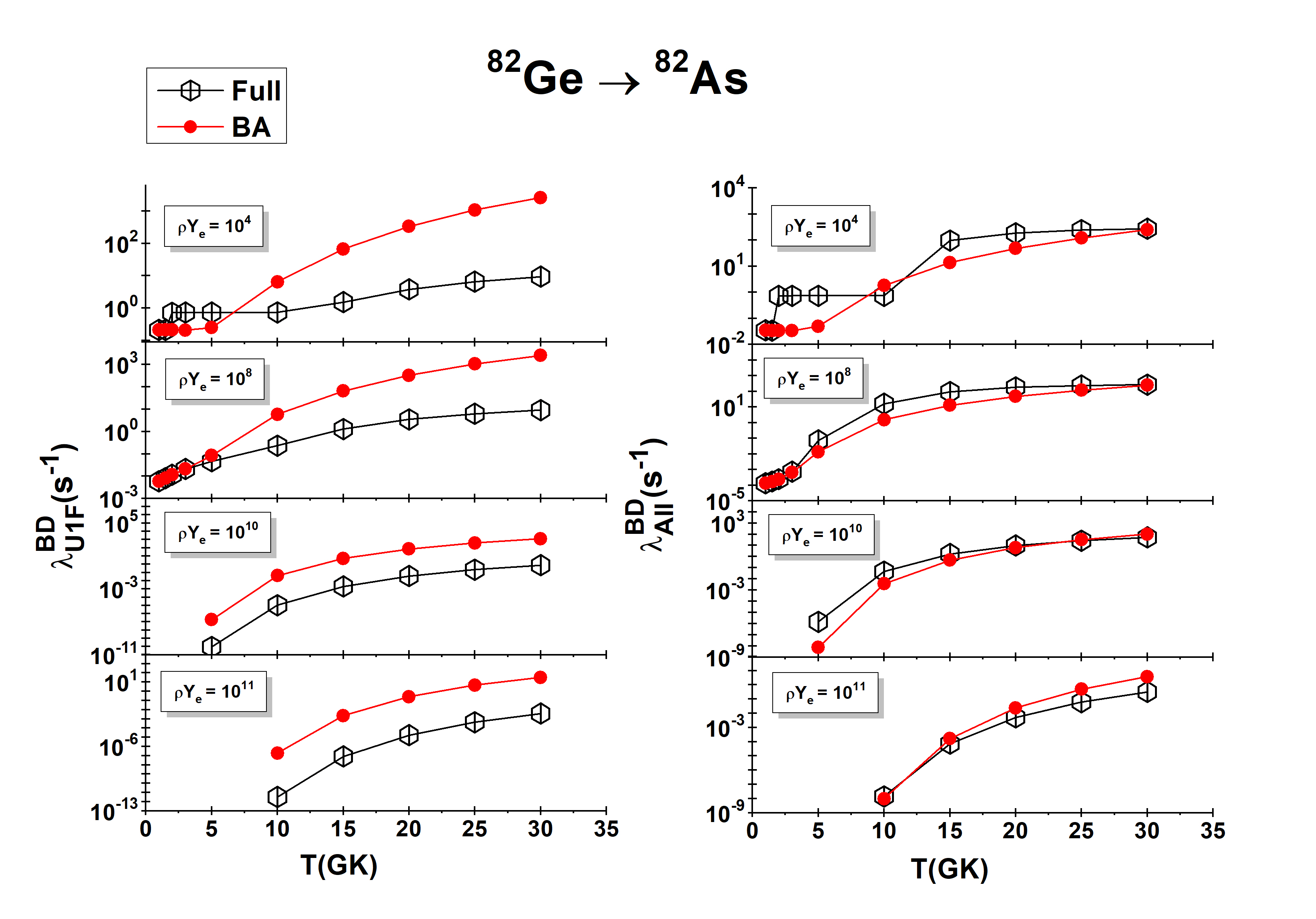}
\end{center}
\caption{\small  Calculated \textit{Full} (microscopic) and \textit{BA} (based on \textit{BA} hypothesis) BD rates [s$^{-1}$] for U1F ($\lambda^{BD}_{U1F}$) and allowed ($\lambda^{BD}_{All}$) transitions  on $^{82}$Ge at selected stellar densities ($\rho \text{Y}_{e}$ [g/cm$^{3}$]) and temperatures.}\label{Fig. 3}
\end{figure}

\begin{figure}[H]
\begin{center}
\includegraphics[width=1.1\textwidth]{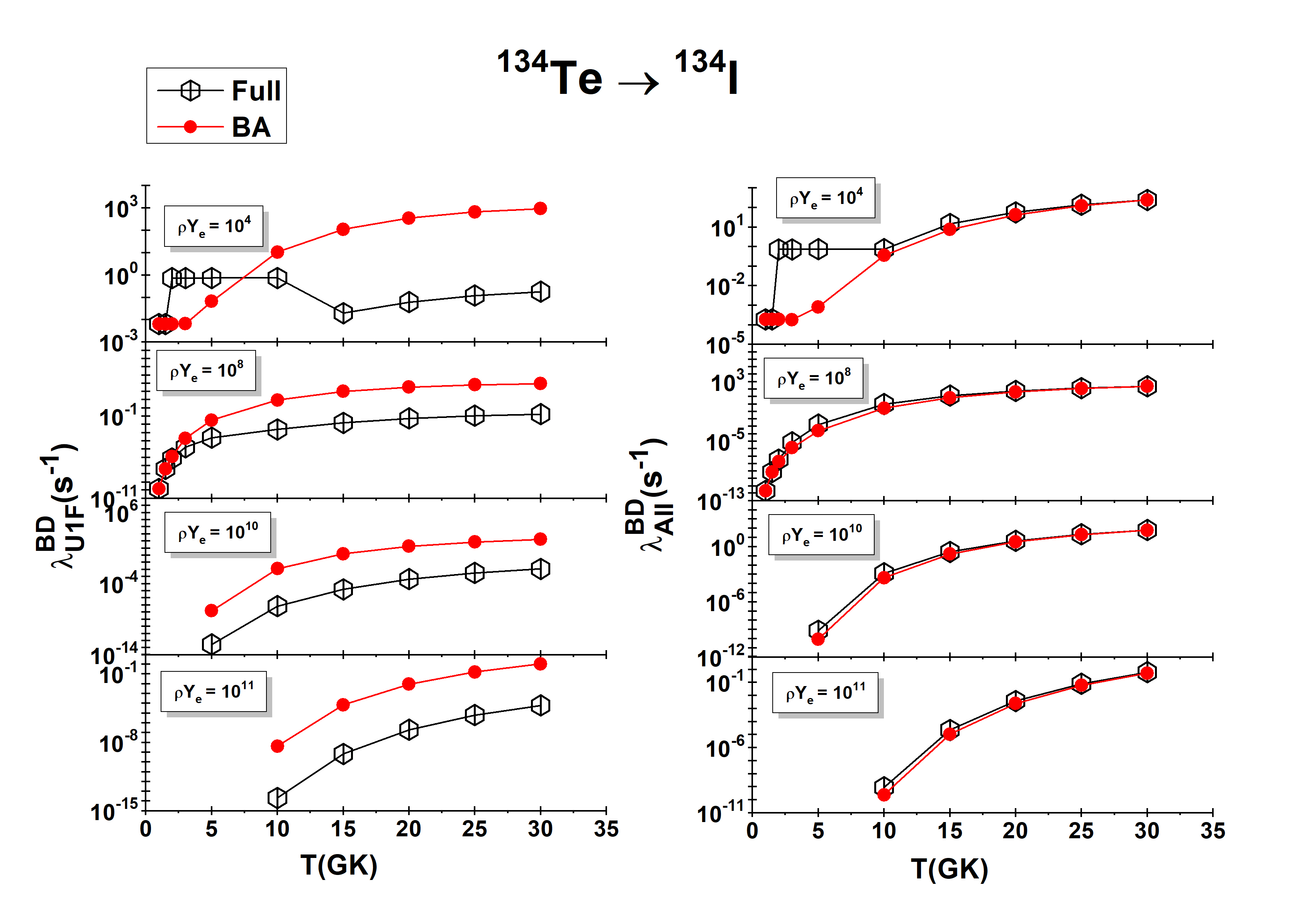}
\end{center}
\caption{\small Same as Fig.~\ref{Fig. 3} but for $^{134}$Te.
}\label{Fig. 4}
\end{figure}

\begin{figure}[H]
\begin{center}
\includegraphics[width=1.1\textwidth]{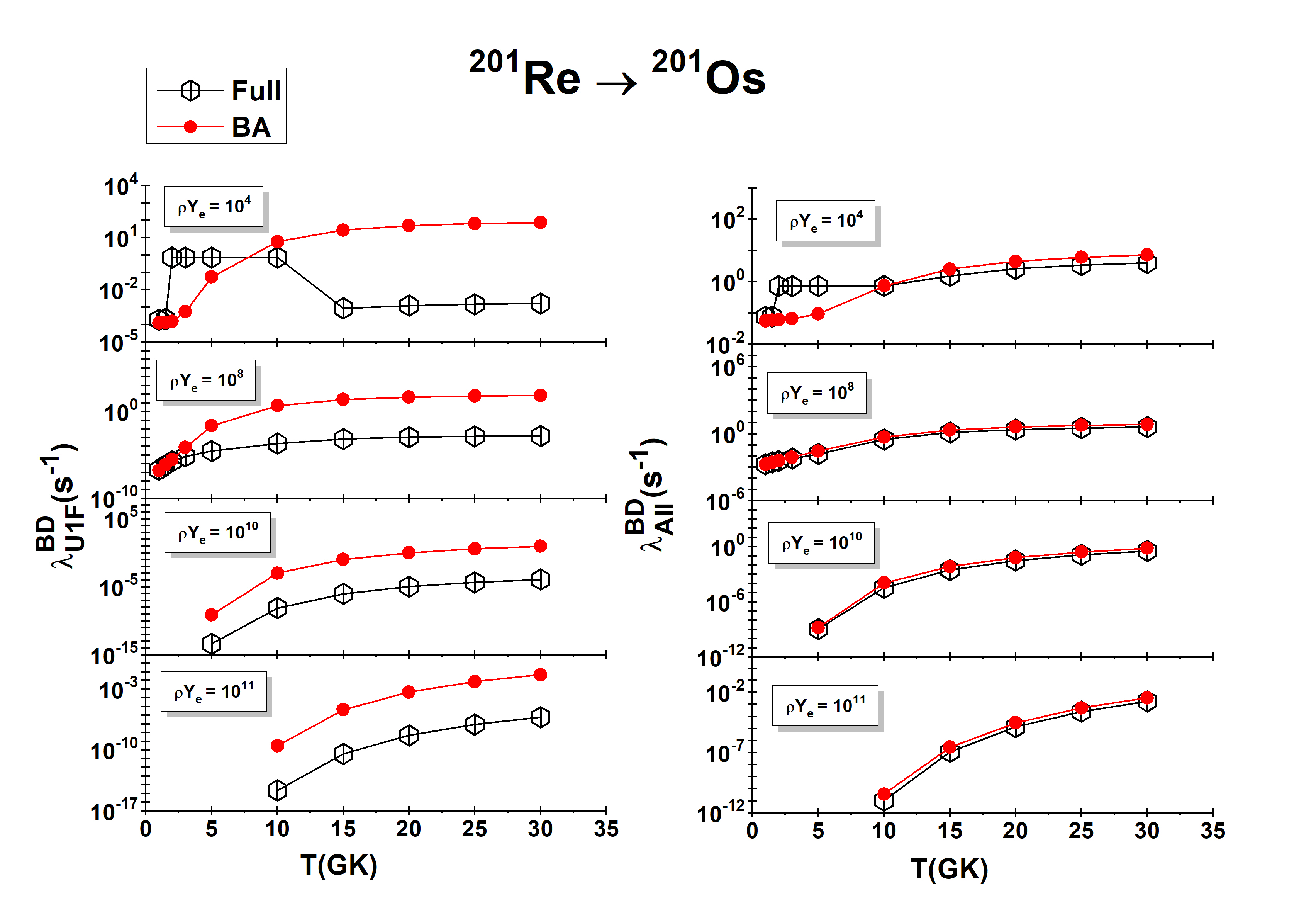}
\end{center}
\caption{\small Same as Fig.~\ref{Fig. 3} but for $^{201}$Re.
}\label{Fig. 5}
\end{figure}

\begin{figure}[H]
\begin{center}
\includegraphics[width=1.1\textwidth]{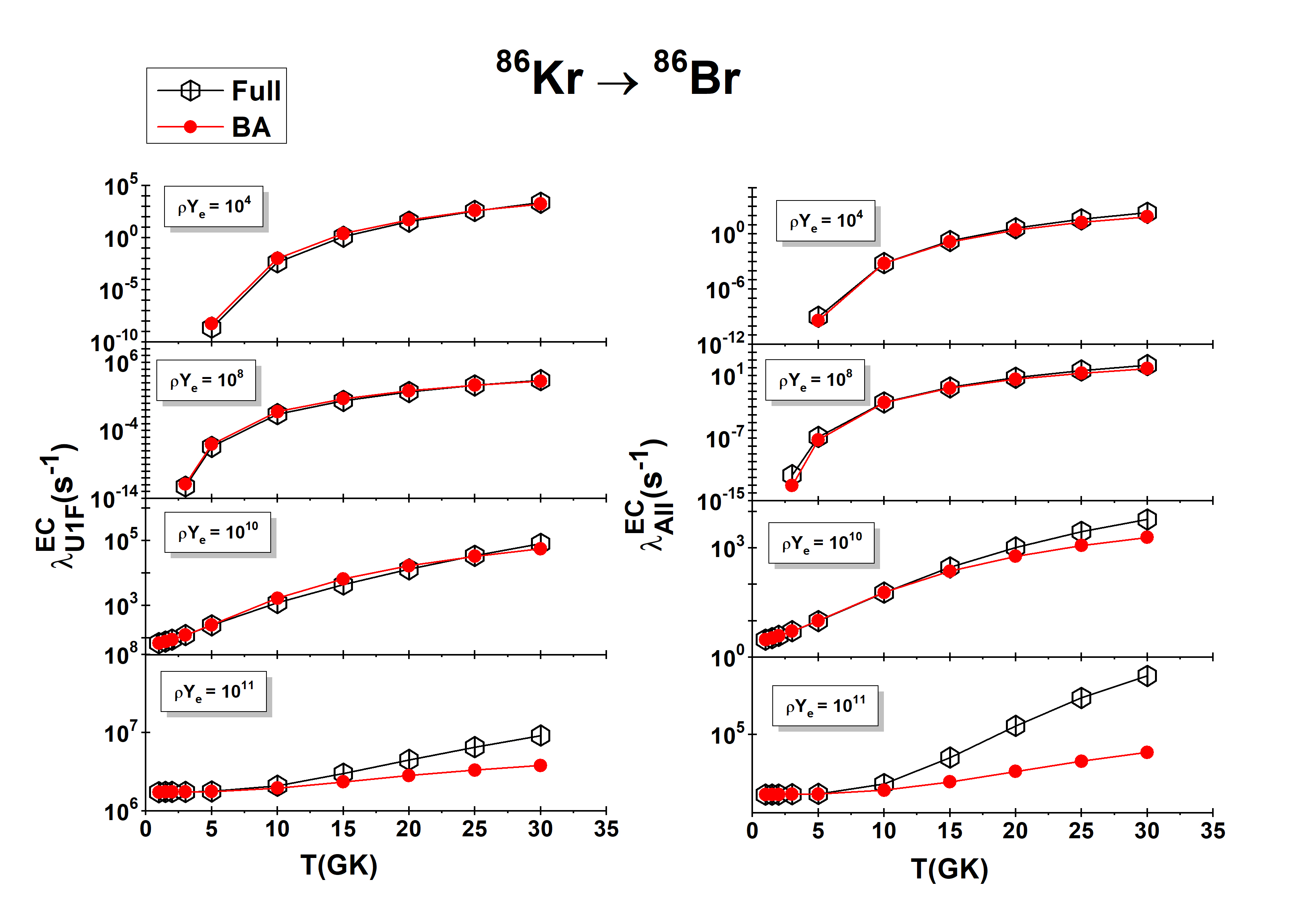}
\end{center}
\caption{\small Same as Fig.~\ref{Fig. 3} but for EC rates of U1F ($\lambda^{EC}_{U1F}$) and allowed ($\lambda^{EC}_{All}$) transitions on $^{86}$Kr.
}\label{Fig. 6}
\end{figure}

\begin{figure}[H]
\begin{center}
\includegraphics[width=1.1\textwidth]{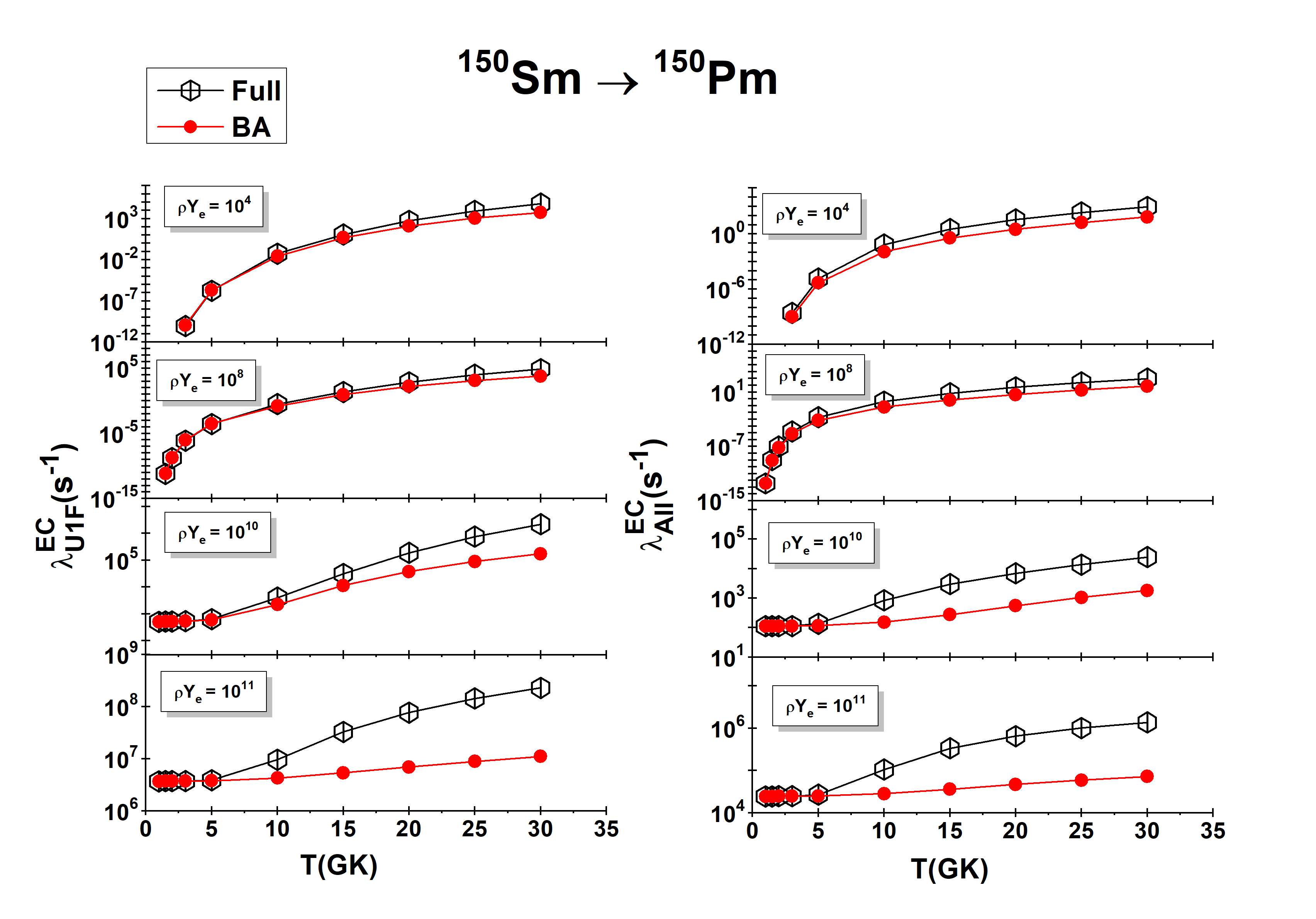}
\end{center}
\caption{\small Same as Fig.~\ref{Fig. 6} but for EC rates on $^{150}$Sm.
}\label{Fig. 7}
\end{figure}

\begin{figure}[H]
\begin{center}
\includegraphics[width=1.1\textwidth]{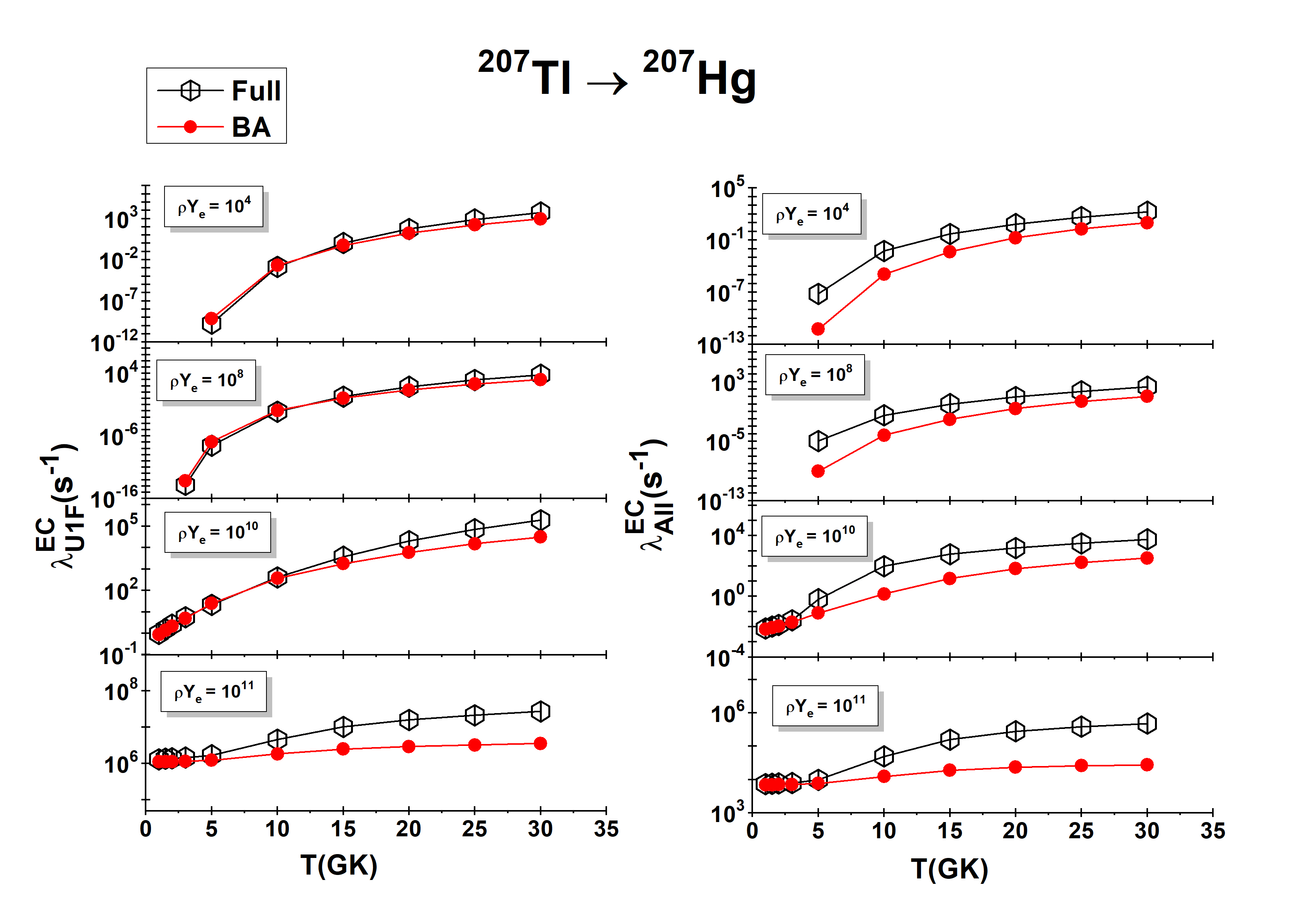}
\end{center}
\caption{\small Same as Fig.~\ref{Fig. 6} but for EC rates on $^{207}$Tl.
}\label{Fig. 8}
\end{figure}

In order to analyze validity of BA hypothesis for calculation of BD rates under stellar conditions, we chose three waiting point nuclei ($^{82}$Ge,  $^{134}$Te and  $^{201}$Re). The selected nuclei have N = 50, N = 82 and N = 126, respectively. Accurate determination of the BD rates of these waiting point nuclei bears significance for the $r$-process nucleosynthesis. Comparison between $\lambda_{F}$ and $\lambda_{BA}$ rates of allowed and U1F transitions is presented in 
Figs.~(\ref{Fig. 3}$-$\ref{Fig. 5}) in BD direction. The effectiveness of applying \textit{BA} hypothesis for EC nuclei is displayed in Figs. 
(\ref{Fig. 6}$-$\ref{Fig. 8}). Here we selected  $^{86}$Kr,  $^{150}$Sm and  $^{207}$Tl as study cases. The values of rates are given in per-second units. In all of these figures, left three panels (in a vertical direction) show both \textit{F(ull)} and \textit{BA} 
U1F rates, whereas, the allowed rates are compared in the right panels. In these figures, $\lambda^{BD}_{All}$ and $\lambda^{BD}_{U1F}$ ($\lambda^{EC}_{All}$ and $\lambda^{EC}_{U1F}$) represent allowed and U1F rates of BD (EC) transitions, respectively. 
In this current study, we have calculated rates for temperature range  T = (1 -- 30) [GK] and density range  $\rho$Y$_{e}$ = (10$^{3}$ -- 10$^{11}$) [g/cm$^{3}$], which roughly corresponds to the physical conditions pertinent to the $r$-process environment. Because of space consideration, 
the results have been reported at selected density snapshots: $\rho$Y$_{e}$ = (10$^{4}$, 10$^{8}$, 10$^{10}$ and 10$^{11}$) [g/cm$^{3}$]. Rates smaller than  10$^{-15}$ [$s^{-1}$] are not shown in the figures. 

A careful analysis of Figs.~(\ref{Fig. 3}$-$\ref{Fig. 5}) shows that 
$\lambda_{BA}$ of U1F BD transitions are much bigger than $\lambda_{F}$. Table \ref{table:table2} shows the ratios (R$_{i}$) between \textit{BA} and \textit{Full} BD rates, calculated according to Eq.~\ref{Ri}, for both allowed GT and U1F transitions, at predetermined physical conditions of the stellar core for three more nuclei. No entries are shown for ratios at core density  $\rho$Y$_{e}$ = 10$^{11}$ [g/cm$^{3}$] as the calculated rates are less than 10$^{-100}$ [s$^{-1}$]. It is noted that at times, the calculated rates are very small ($\lambda$ $<$ 10$^{-5}$ [s$^{-1}$]). These very small
numbers can change by orders of magnitude by a mere
change of 0.5 [MeV], or less, in parent or daughter excitation
energies and are more reflective of the uncertainties
in the calculation of energies \cite{Nab04}. Consequently we show two different values of average deviation towards the end of Table \ref{table:table2}. The first entry is the one defined by Eq.~\ref{Rbar}. The second entry (marked with an *) is the one excluding entries where calculated rates are less than 10$^{-5}$ s$^{-1}$.   Table \ref{table:table3} depicts similar data for EC rates.  Our results show that the U1F BD rates, calculated by incorporating the 
\textit{BA} hypothesis, are largely overestimated relative to the microscopic (\textit{Full}) rates by up to 4--5 orders of magnitude. On the other hand, Tables~(\ref{table:table2} -- \ref{table:table3}) show that for EC (U1F and allowed) and allowed BD, the \textit{BA} rates are, on the average, smaller than the \textit{Full} rates. 
The total (allowed plus U1F) \textit{BA} rates ($\lambda_{BA} [All+U1F]$), in both EC and BD directions, deviate from total \textit{Full} ($\lambda_{F} [All+U1F]$) rates, on the average, by an order of magnitude or more. This difference can be seen  from Table~\ref{table:table4} and Table~\ref{table:table5} for BD and EC, respectively. These tables show total \textit{BA} and \textit{Full} rates and ratios (R$_{i}$) 
between them.

\begin{table}[h!]
\caption{\small Comparison of \textit{BA} ($\lambda_{BA}$) and \textit{Full} ($\lambda_{F}$) $\beta$-decay rates for three selected nuclei as a function of core temperature (T [GK]) and density ($\rho \text{Y}_{e}$  [g/cm$^{3}$]). Ratios (R$_{i}$) and average deviation ($\bar{R}$) are defined in Eq.~\ref{Ri} and Eq.~\ref{Rbar}, respectively. Computed R$_{i}$ values, where $\lambda_{F}$ and/or $\lambda_{BA}$ rates are less than 10$^{-5}$ s$^{-1}$, are marked with  *. $\bar{R}^{(*)}$ are computed average deviations excluding ratios marked with *.}\label{table:table2}\setlength{\tabcolsep}{2pt}
\centering {\small    
\resizebox{17cm}{!}{%
\begin{tabular}{cc|c|c|c|c|c|c|c|c|c|c|c|c}
\toprule \multirow{3}{*}{\textbf{T}} &
				\multirow{3}{*}{$\mathbf{\brho Y_{e}}$} & \multicolumn{4}{c|}{\textbf{$^{82}$Ge}}&
				\multicolumn{4}{c|}{\textbf{$^{134}$Te}} & \multicolumn{4}{c}{\textbf{$^{201}$Re}}\\
				\cmidrule{3-6}  \cmidrule{7-10}  \cmidrule{11-14} & & 
				\multicolumn{2}{|c|}{\bf{R$_{i}$} [All]} &
				\multicolumn{2}{c|}{\bf{R$_{i}$} [U1F]} & 
				\multicolumn{2}{c|}{\bf{R$_{i}$} [All]} &
				\multicolumn{2}{c|}{\bf{R$_{i}$} [U1F]} & 
				\multicolumn{2}{c|}{\bf{R$_{i}$} [All]} &
				\multicolumn{2}{c}{\bf{R$_{i}$} [U1F]} \\ & & 
				\multicolumn{1}{c|}{\bf{$\lambda_{F} \ge \lambda_{BA}$}} &
				\multicolumn{1}{c|}{\bf{$\lambda_{BA}>\lambda_{F}$}} &
				\multicolumn{1}{c|}{\bf{$\lambda_{F} \ge \lambda_{BA}$}} &
				\multicolumn{1}{c|}{\bf{$\lambda_{BA}>\lambda_{F}$}}  & 
				\multicolumn{1}{c|}{\bf{$\lambda_{F} \ge \lambda_{BA}$}} &
				\multicolumn{1}{c|}{\bf{$\lambda_{BA}>\lambda_{F}$}} & 
				\multicolumn{1}{c|}{\bf{$\lambda_{F} \ge \lambda_{BA}$}} &
				\multicolumn{1}{c|}{\bf{$\lambda_{BA}>\lambda_{F}$}} &
				\multicolumn{1}{c|}{\bf{$\lambda_{F} \ge \lambda_{BA}$}}  & 
				\multicolumn{1}{c|}{\bf{$\lambda_{BA}>\lambda_{F}$}} &
				\multicolumn{1}{c|}{\bf{$\lambda_{F} \ge \lambda_{BA}$}}  & 
				\multicolumn{1}{c}{\bf{$\lambda_{BA}>\lambda_{F}$}} \\
				\hline
				1     & 10$^{4}$ & 1.0   &       &       & 1.0   & 1.0   &       &       & 1.0   & 1.3   &       & 1.5   &  \\
				5     & 10$^{4}$ & 2.5   &       &       & 1.2   & 3.7   &       &       & 12.2  &       & 1.2   &       & 264.9 \\
				10    & 10$^{4}$ & 9.9   &       &       & 17.3  & 3.3   &       &       & 1840.8 &       & 1.5   &       & 15452.5 \\
				20    & 10$^{4}$ & 3.8   &       &       & 86.7  & 1.4   &       &       & 5821.0 &       & 1.7   &       & 39355.0 \\
				30    & 10$^{4}$ & 1.1   &       &       & 272.3 & 1.0   &       &       & 5164.2 &       & 1.8   &       & 46025.7 \\
				&       &       &       &       &       &       &       &       &       &       &       &       &  \\
				1     & 10$^{8}$ & 1.0   &       &       & 1.0   & 1.0$^{(*)}$   &       &       & 1.0$^{(*)}$   & 1.1   &       & 1.1   &  \\
				5     & 10$^{8}$ & 5.3   &       &       & 1.8   & 6.5   &       &       & 139.0 &       & 1.8   &       & 824.1 \\
				10    & 10$^{8}$ & 10.7  &       &       & 25.2  & 3.6   &       &       & 3698.3 &       & 1.6   &       & 22335.7 \\
				20    & 10$^{8}$ & 3.8   &       &       & 93.1  & 1.4   &       &       & 6295.1 &       & 1.7   &       & 41115.0 \\
				30    & 10$^{8}$ & 1.1   &       &       & 279.3 & 1.0   &       &       & 5284.5 &       & 1.8   &       & 46558.6 \\
				&       &       &       &       &       &       &       &       &       &       &       &       &  \\
				1     & 10$^{11}$ & ---     & ---     & ---     & ---     & ---     & ---     & ---     & ---     & ---     & ---     & ---     & --- \\
				5     & 10$^{11}$ & 24.1$^{(*)}$  &       &       & 40457.6$^{(*)}$ & 6.1$^{(*)}$   &       &       & 53210.8$^{(*)}$ &       & 1.5$^{(*)}$   &       & 17418.1$^{(*)}$ \\
				10    & 10$^{11}$ & 1.4$^{(*)}$   &       &       & 58210.3$^{(*)}$ & 3.8$^{(*)}$   &       &       & 199067.3$^{(*)}$ &       & 3.4$^{(*)}$   &       & 133967.7$^{(*)}$ \\
				20    & 10$^{11}$ &       & 4.7   &       & 15417.0 & 1.6   &       &       & 45081.7$^{(*)}$ &       & 2.1   &       & 97949.0$^{(*)}$ \\
				30    & 10$^{11}$ &       & 12.0  &       & 8609.9 & 1.2   &       &       & 17458.2 &       & 2.0   &       & 78343.0$^{(*)}$ \\
				\hline
				\multicolumn{2}{c|}{$\mathbf{\bar{R}}$} &       \multicolumn{2}{c|}{5.9} & \multicolumn{2}{c|}{8819.6} & \multicolumn{2}{c|}{2.6} & \multicolumn{2}{c|}{24505.4} & \multicolumn{2}{c|}{1.7} & \multicolumn{2}{c}{38543.7}\\
				\hline
				\multicolumn{2}{c|}{$\mathbf{\bar{R}^{(*)}}$} &       \multicolumn{2}{c|}{4.7} & \multicolumn{2}{c|}{2067.2} & \multicolumn{2}{c|}{2.3} & \multicolumn{2}{c|}{4571.4} & \multicolumn{2}{c|}{1.6} & \multicolumn{2}{c}{23548.1} \\
	\end{tabular}}}
\end{table}
\begin{table}[h!]
	\caption{\small Same as Table~\ref{table:table2} but for three $r$-process EC nuclei.
	}\label{table:table3} 
	\setlength{\tabcolsep}{2pt}
	\centering {\small
		\resizebox{17cm}{!}{%
			\begin{tabular}{cc|c|c|c|c|c|c|c|c|c|c|c|c}
				\toprule \multirow{3}{*}{\textbf{T}} &
				\multirow{3}{*}{$\mathbf{\brho Y_{e}}$} & \multicolumn{4}{|c|}{\textbf{$^{86}$Kr}}&
				\multicolumn{4}{c|}{\textbf{$^{150}$Sm}} & \multicolumn{4}{c}{\textbf{$^{207}$Tl}}\\
				\cmidrule{3-6}  \cmidrule{7-10}  \cmidrule{11-14} & & 
				\multicolumn{2}{|c|}{\bf{R$_{i}$} [All]} &
				\multicolumn{2}{c|}{\bf{R$_{i}$} [U1F]} & 
				\multicolumn{2}{c|}{\bf{R$_{i}$} [All]} &
				\multicolumn{2}{c|}{\bf{R$_{i}$} [U1F]} & 
				\multicolumn{2}{c|}{\bf{R$_{i}$} [All]} &
				\multicolumn{2}{c}{\bf{R$_{i}$} [U1F]} \\ & & 
				\multicolumn{1}{c|}{\bf{$\lambda_{F}\ge \lambda_{BA}$}} &
				\multicolumn{1}{c|}{\bf{$\lambda_{BA}>\lambda_{F}$}} &
				\multicolumn{1}{c|}{\bf{$\lambda_{F} \ge \lambda_{BA}$}} &
				\multicolumn{1}{c|}{\bf{$\lambda_{BA}>\lambda_{F}$}}  & 
				\multicolumn{1}{c|}{\bf{$\lambda_{F} \ge \lambda_{BA}$}} &
				\multicolumn{1}{c|}{\bf{$\lambda_{BA}>\lambda_{F}$}} & 
				\multicolumn{1}{c|}{\bf{$\lambda_{F} \ge \lambda_{BA}$}} &
				\multicolumn{1}{c|}{\bf{$\lambda_{BA}>\lambda_{F}$}} &
				\multicolumn{1}{c|}{\bf{$\lambda_{F} \ge \lambda_{BA}$}}  & 
				\multicolumn{1}{c|}{\bf{$\lambda_{BA}>\lambda_{F}$}} &
				\multicolumn{1}{c|}{\bf{$\lambda_{F} \ge \lambda_{BA}$}}  & 
				\multicolumn{1}{c}{\bf{$\lambda_{BA}>\lambda_{F}$}} \\
				\hline
				1     & 10$^{4}$ & 1.2$\times$10$^{+08(*)}$ &       &       & 2.1$^{(*)}$  & 3.2$^{(*)}$  &       &       & 1.2$^{(*)}$  & 5.0$\times$10$^{+21(*)}$ &       &       & 6.0$^{(*)}$ \\
				5     & 10$^{4}$ & 2.5$^{(*)}$  &       &       & 2.4$^{(*)}$  & 2.9$^{(*)}$  &       &       & 1.2$^{(*)}$  & 1.2$\times$10$^{+04(*)}$ &       &       & 4.1$^{(*)}$ \\
				10    & 10$^{4}$ & 1.1  &       &       & 2.4  & 5.9  &       & 2.0  &       & 485.3 &       &       & 1.5 \\
				20    & 10$^{4}$ & 1.6  &       &       & 1.5  & 11.6 &       & 4.3  &       & 37.5 &       & 3.2  &     \\
				30    & 10$^{4}$ & 2.8  &       & 1.3  &       & 12.7 &       & 11.5 &       & 19.3 &       & 6.0  &     \\
				&       &       &       &       &       &       &       &       &       &       &       &       &  \\
				1     & 10$^{8}$ & 5.1$\times$10$^{+07(*)}$ &       &       & 2.1 $^{(*)}$ & 1.0 $^{(*)}$ &       &       & 1.1$^{(*)}$  & 4.7$\times$10$^{+21(*)}$ &       &       & 6.0$^{(*)}$ \\
				5     & 10$^{8}$ & 2.4$^{(*)}$  &       &       & 2.4$^{(*)}$  & 2.9  &       &       & 1.2  & 1.2$\times$10$^{+04(*)}$ &       &       & 4.1$^{(*)}$ \\
				10    & 10$^{8}$ & 1.1  &       &       & 2.4  & 6.0  &       & 2.0  &       & 479.7 &       &       & 1.5 \\
				20    & 10$^{8}$ & 1.6  &       &       & 1.5  & 11.6 &       & 4.3  &       & 37.4 &       & 3.2  &     \\
				30    & 10$^{8}$ & 2.8  &       & 1.3  &       & 12.7 &       & 11.5 &       & 19.3 &       & 6.0  &     \\
				&       &       &       &       &       &       &       &       &       &       &       &       &  \\
				1     & 10$^{11}$ & 1.0  &       & 1.0  &       & 1.0  &       & 1.0  &       & 1.1  &       & 1.2  &  \\
				5     & 10$^{11}$ & 1.0  &       & 1.0  &       & 1.1  &       & 1.0  &       & 1.3  &       & 1.4  &  \\
				10    & 10$^{11}$ & 1.1  &       & 1.1  &       & 3.7  &       & 2.2  &       & 4.0  &       & 2.5  &  \\
				20    & 10$^{11}$ & 2.6  &       & 1.6  &       & 13.9 &       & 11.1 &       & 12.0 &       & 5.4  &  \\
				30    & 10$^{11}$ & 4.8  &       & 2.4  &       & 18.9 &       & 20.7 &       & 17.5 &       & 7.6  &  \\
				\hline
				\multicolumn{2}{c|}{$\mathbf{\bar{R}}$} & \multicolumn{2}{c|}{1.1$\times$10$^{+07}$} & \multicolumn{2}{c|}{1.8} & \multicolumn{2}{c|}{7.3} & \multicolumn{2}{c|}{5.1} & \multicolumn{2}{c|}{6.4$\times$10$^{+20}$} & \multicolumn{2}{c}{4.0} 
				\\ 
				\hline
				\multicolumn{2}{c|}{$\mathbf{\bar{R}^{(*)}}$} &    \multicolumn{2}{c|}{2.0} & \multicolumn{2}{c|}{1.6} & \multicolumn{2}{c|}{8.5} & \multicolumn{2}{c|}{6.1} & \multicolumn{2}{c|}{101.3} & \multicolumn{2}{c}{3.6} 
					\end{tabular}}}
\end{table}
\begin{table}[h!]
\caption{\small Comparison of total \textit{BA} ($\lambda_{BA} [All+U1F]$) and total \textit{Full} ($\lambda_{F} [All+U1F]$) $\beta$-decay rates for three selected nuclei as a function of core temperature (T [GK]) and density ($\rho \text{Y}_{e}$  [g/cm$^{3}$]). Ratios (R$_{i}$) and average deviation ($\bar{R}$)  are defined in Eq.~\ref{Ri} and Eq.~\ref{Rbar}, respectively. Computed R$_{i}$ values, where $\lambda_{F}$ $>$ $\lambda_{BA}$, are marked with  *.}\label{table:table4}\setlength{\tabcolsep}{2pt}
\centering {\small    
\resizebox{17cm}{!}{%
\begin{tabular}{cc|c|c|c|c|c|c|c|c|c}
\toprule \multirow{2}{*}{\textbf{T}} &
				\multirow{2}{*}{$\mathbf{\brho Y_{e}}$} & \multicolumn{3}{c|}{\textbf{$^{82}$Ge}}&
				\multicolumn{3}{c|}{\textbf{$^{134}$Te}} & \multicolumn{3}{c}{\textbf{$^{201}$Re}}\\
				\cmidrule{3-5}  \cmidrule{6-8}  \cmidrule{9-11}  & & 
				\multicolumn{1}{|c|}{$\mathbf{\lambda_{F} [All+U1F]}$} &
				\multicolumn{1}{c|}{$\mathbf{\lambda_{BA} [All+U1F]}$} & 
				\multicolumn{1}{c|}{\bf{R$_{i}$ ($\lambda_{BA} \ge \lambda_{F}$)}} &
				\multicolumn{1}{|c|}{$\mathbf{\lambda_{F} [All+U1F]}$} &
				\multicolumn{1}{c|}{$\mathbf{\lambda_{BA} [All+U1F]}$} & 
				\multicolumn{1}{c|}{\bf{R$_{i}$ ($\lambda_{BA} \ge \lambda_{F}$)}} &
				\multicolumn{1}{|c|}{$\mathbf{\lambda_{F} [All+U1F]}$} &
				\multicolumn{1}{c|}{$\mathbf{\lambda_{BA} [All+U1F]}$} & 
				\multicolumn{1}{c}{\bf{R$_{i}$ ($\lambda_{BA} \ge \lambda_{F}$)}}  \\ 
				\hline
    1     & 10$^{4}$ & 2.47$\times$10$^{-01}$ & 2.47$\times$10$^{-01}$ & 1.0   & 6.66$\times$10$^{-03}$ & 6.66$\times$10$^{-03}$ & 1.0   & 7.61$\times$10$^{-02}$ & 5.71$\times$10$^{-02}$ & 1.3$^{(*)}$ \\
    5     & 10$^{4}$ & 3.29$\times$10$^{-01}$ & 3.02$\times$10$^{-01}$ & 1.1$^{(*)}$   & 8.51$\times$10$^{-03}$ & 6.80$\times$10$^{-02}$ & 8.0   & 7.75$\times$10$^{-02}$ & 1.49$\times$10$^{-01}$ & 1.9 \\
    10    & 10$^{4}$ & 1.85$\times$10$^{+01}$ & 8.33$\times$10$^{+00}$ & 2.2$^{(*)}$   & 1.18$\times$10$^{+00}$ & 1.10$\times$10$^{+01}$ & 9.3   & 5.05$\times$10$^{-01}$ & 6.78$\times$10$^{+00}$ & 13.4 \\
    20    & 10$^{4}$ & 1.89$\times$10$^{+02}$ & 3.79$\times$10$^{+02}$ & 2.0   & 5.76$\times$10$^{+01}$ & 3.98$\times$10$^{+02}$ & 6.9   & 2.62$\times$10$^{+00}$ & 5.45$\times$10$^{+01}$ & 20.8 \\
    30    & 10$^{4}$ & 2.82$\times$10$^{+02}$ & 2.80$\times$10$^{+03}$ & 10.0  & 2.47$\times$10$^{+02}$ & 1.18$\times$10$^{+03}$ & 4.8   & 4.06$\times$10$^{+00}$ & 8.31$\times$10$^{+01}$ & 20.5 \\
          &       &       &       &       &       &       &       &       &       &  \\
    1     & 10$^{8}$ & 6.02$\times$10$^{-03}$ & 6.02$\times$10$^{-03}$ & 1.0   & 1.48$\times$10$^{-10}$ & 1.48$\times$10$^{-10}$ & 1.0   & 1.96$\times$10$^{-03}$ & 1.83$\times$10$^{-03}$ & 1.1$^{(*)}$ \\
    5     & 10$^{8}$ & 1.19$\times$10$^{-01}$ & 9.85$\times$10$^{-02}$ & 1.2$^{(*)}$   & 1.94$\times$10$^{-03}$ & 3.44$\times$10$^{-02}$ & 17.8  & 1.57$\times$10$^{-02}$ & 5.31$\times$10$^{-02}$ & 3.4 \\
    10    & 10$^{8}$ & 1.65$\times$10$^{+01}$ & 7.41$\times$10$^{+00}$ & 2.2$^{(*)}$  & 1.00$\times$10$^{+00}$ & 9.63$\times$10$^{+00}$ & 9.6   & 3.18$\times$10$^{-01}$ & 5.27$\times$10$^{+00}$ & 16.6 \\
    20    & 10$^{8}$ & 1.83$\times$10$^{+02}$ & 3.73$\times$10$^{+02}$ & 2.0   & 5.58$\times$10$^{+01}$ & 3.87$\times$10$^{+02}$ & 6.9   & 2.42$\times$10$^{+00}$ & 5.17$\times$10$^{+01}$ & 21.3 \\
    30    & 10$^{8}$ & 2.78$\times$10$^{+02}$ & 2.79$\times$10$^{+03}$ & 10.1  & 2.44$\times$10$^{+02}$ & 1.17$\times$10$^{+03}$ & 4.8   & 3.95$\times$10$^{+00}$ & 8.14$\times$10$^{+01}$ & 20.6 \\
				&       &       &       &       &       &       &       &       &       &       \\
1     & 10$^{11}$ & ---     & ---     & ---     & ---     & ---     & ---     & ---     & ---     & --- \\
	 5     & 10$^{11}$ & 1.81$\times$10$^{-19}$ & 4.15$\times$10$^{-19}$ & 2.3   & 1.98$\times$10$^{-22}$ & 1.79$\times$10$^{-21}$ & 9.1   & 1.26$\times$10$^{-22}$ & 2.70$\times$10$^{-22}$ & 2.1 \\
    10    & 10$^{11}$ & 1.44$\times$10$^{-08}$ & 2.00$\times$10$^{-07}$ & 13.9  & 8.95$\times$10$^{-10}$ & 4.37$\times$10$^{-09}$ & 4.9   & 1.05$\times$10$^{-11}$ & 3.39$\times$10$^{-10}$ & 32.2 \\
    20    & 10$^{11}$ & 5.05$\times$10$^{-03}$ & 2.83$\times$10$^{-01}$ & 55.9  & 3.78$\times$10$^{-03}$ & 1.14$\times$10$^{-02}$ & 3.0   & 1.41$\times$10$^{-05}$ & 4.97$\times$10$^{-04}$ & 35.2 \\
    30    & 10$^{11}$ & 3.19$\times$10$^{-01}$ & 3.58$\times$10$^{+01}$ & 112.1 & 6.01$\times$10$^{-01}$ & 1.57$\times$10$^{+00}$ & 2.6   & 1.79$\times$10$^{-03}$ & 5.04$\times$10$^{-02}$ & 28.1 \\
\hline
\multicolumn{2}{c|}{$\mathbf{\bar{R}}$}   &       &   &15.5  &       &       & 6.4   &       &       & 15.6 \\
\end{tabular}}}
\end{table}
\begin{table}[h!]
\caption{\small Same as Table~\ref{table:table4} but for three $r$-process EC nuclei. Computed R$_{i}$ values, where $\lambda_{BA} > \lambda_{F}$, are marked with  *.}\label{table:table5}\setlength{\tabcolsep}{2pt}
\centering {\small    
\resizebox{17cm}{!}{%
\begin{tabular}{cc|c|c|c|c|c|c|c|c|c}
\toprule \multirow{2}{*}{\textbf{T}} &
				\multirow{2}{*}{$\mathbf{\brho Y_{e}}$} & \multicolumn{3}{|c|}{\textbf{$^{86}$Kr}}&
				\multicolumn{3}{c|}{\textbf{$^{150}$Sm}} & \multicolumn{3}{c}{\textbf{$^{207}$Tl}}\\
				\cmidrule{3-5}  \cmidrule{6-8}  \cmidrule{9-11}  & & 
				\multicolumn{1}{|c|}{$\mathbf{\lambda_{F} [All+U1F]}$} &
				\multicolumn{1}{c|}{$\mathbf{\lambda_{BA} [All+U1F]}$} & 
				\multicolumn{1}{c|}{\bf{R$_{i}$ ($\lambda_{F} \ge \lambda_{BA}$)}} &
				\multicolumn{1}{|c|}{$\mathbf{\lambda_{F} [All+U1F]}$} &
				\multicolumn{1}{c|}{$\mathbf{\lambda_{BA} [All+U1F]}$} & 
				\multicolumn{1}{c|}{\bf{R$_{i}$ ($\lambda_{F} \ge \lambda_{BA}$)}} &
				\multicolumn{1}{|c|}{$\mathbf{\lambda_{F} [All+U1F]}$} &
				\multicolumn{1}{c|}{$\mathbf{\lambda_{BA} [All+U1F]}$} & 
				\multicolumn{1}{c}{\bf{R$_{i}$ ($\lambda_{F} \ge \lambda_{BA}$)}}  \\ 
				\hline
    1     & 10$^{4}$ &  ---     & ---     & ---     & ---     & ---     & ---     & ---     & ---     & --- \\
    5     & 10$^{4}$ & 3.49$\times$10$^{-09}$ & 6.33$\times$10$^{-09}$ & 1.8$^{(*)}$   & 1.63$\times$10$^{-05}$ & 7.13$\times$10$^{-06}$ & 2.3   & 6.81$\times$10$^{-08}$ & 7.30$\times$10$^{-10}$ & 93.3 \\
    10    & 10$^{4}$ & 5.10$\times$10$^{-03}$ & 1.11$\times$10$^{-02}$ & 2.2$^{(*)}$   & 1.27$\times$10$^{-01}$ & 3.96$\times$10$^{-02}$ & 3.2   & 7.00$\times$10$^{-03}$ & 2.06$\times$10$^{-03}$ & 3.4 \\
    20    & 10$^{4}$ & 4.04$\times$10$^{+01}$ & 5.60$\times$10$^{+01}$ & 1.4$^{(*)}$   & 5.75$\times$10$^{+02}$ & 1.28$\times$10$^{+02}$ & 4.5   & 6.16$\times$10$^{+01}$ & 1.70$\times$10$^{+01}$ & 3.6 \\
    30    & 10$^{4}$ & 2.55$\times$10$^{+03}$ & 1.81$\times$10$^{+03}$ & 1.4   & 6.53$\times$10$^{+04}$ & 5.67$\times$10$^{+03}$ & 11.5  & 5.72$\times$10$^{+03}$ & 9.30$\times$10$^{+02}$ & 6.2 \\
          &       &       &       &       &       &       &       &       &       &  \\
    1     & 10$^{8}$ &  ---     & ---     & ---     & ---     & ---     & ---     & ---     & ---     & --- \\
    5     & 10$^{8}$ & 5.50$\times$10$^{-07}$ & 1.02$\times$10$^{-06}$ & 1.9$^{(*)}$   & 2.28$\times$10$^{-03}$ & 1.03$\times$10$^{-03}$ & 2.2   & 1.07$\times$10$^{-05}$ & 1.18$\times$10$^{-07}$ & 90.9 \\
    10    & 10$^{8}$ & 2.88$\times$10$^{-02}$ & 6.28$\times$10$^{-02}$ & 2.2$^{(*)}$   & 7.09$\times$10$^{-01}$ & 2.20$\times$10$^{-01}$ & 3.2   & 3.92$\times$10$^{-02}$ & 1.16$\times$10$^{-02}$ & 3.4 \\
    20    & 10$^{8}$ & 5.31$\times$10$^{+01}$ & 7.34$\times$10$^{+01}$ & 1.4$^{(*)}$   & 7.56$\times$10$^{+02}$ & 1.68$\times$10$^{+02}$ & 4.5   & 8.10$\times$10$^{+01}$ & 2.24$\times$10$^{+01}$ & 3.6 \\
    30    & 10$^{8}$ & 2.77$\times$10$^{+03}$ & 1.96$\times$10$^{+03}$ & 1.4   & 7.08$\times$10$^{+04}$ & 6.14$\times$10$^{+03}$ & 11.5  & 6.20$\times$10$^{+03}$ & 1.01$\times$10$^{+03}$ & 6.2 \\
				&       &       &       &       &       &       &       &       &       &       \\
    1     & 10$^{11}$ & 1.77$\times$10$^{+06}$ & 1.77$\times$10$^{+06}$ & 1.0   & 3.72$\times$10$^{+06}$ & 3.72$\times$10$^{+06}$ & 1.0   & 1.32$\times$10$^{+06}$ & 1.13$\times$10$^{+06}$ & 1.2 \\
    5     & 10$^{11}$ & 1.80$\times$10$^{+06}$ & 1.80$\times$10$^{+06}$ & 1.0   & 3.92$\times$10$^{+06}$ & 3.81$\times$10$^{+06}$ & 1.0   & 1.70$\times$10$^{+06}$ & 1.24$\times$10$^{+06}$ & 1.4 \\
    10    & 10$^{11}$ & 2.14$\times$10$^{+06}$ & 1.99$\times$10$^{+06}$ & 1.1   & 9.70$\times$10$^{+06}$ & 4.29$\times$10$^{+06}$ & 2.3   & 4.61$\times$10$^{+06}$ & 1.84$\times$10$^{+06}$ & 2.5 \\
    20    & 10$^{11}$ & 4.61$\times$10$^{+06}$ & 2.88$\times$10$^{+06}$ & 1.6   & 7.77$\times$10$^{+07}$ & 6.98$\times$10$^{+06}$ & 11.1  & 1.61$\times$10$^{+07}$ & 2.94$\times$10$^{+06}$ & 5.5 \\
    30    & 10$^{11}$ & 9.50$\times$10$^{+06}$ & 3.87$\times$10$^{+06}$ & 2.5   & 2.31$\times$10$^{+08}$ & 1.12$\times$10$^{+07}$ & 20.7  & 2.77$\times$10$^{+07}$ & 3.60$\times$10$^{+06}$ & 7.7 \\
\hline
\multicolumn{2}{c|}{$\mathbf{\bar{R}}$}   &       &       &1.6   &       &       & 6.1   &       &       & 17.6 \\
\end{tabular}}}
\end{table}

There are three main causes of an increase in the calculated weak rate: (1) enlarged phase space,  (2)  bigger total GT strength distribution values, and (3) lower placement of computed GT centroid. The calculated phase space and total GT strength distributions (along with centroid placement) are shown in Tables (\ref{table:table6} -- \ref{table:table7}) and Table~(\ref{table:table8}), respectively. The enhancement of $\lambda_{BA}$ relative to the $\lambda_{F}$, in U1F BD case, is due to enlarged available phase spaces introduced by applying the \textit{BA} hypothesis. Table~\ref{table:table6} shows that, for high core temperatures,  the U1F computed phase spaces by invoking \textit{BA} hypothesis are up to 3 orders of magnitude bigger. For low temperature and high density regions, the calculated BD rates approach zero because of choked phase spaces. In contrast, for allowed GT rates in BD direction, and for both U1F and allowed rates in EC direction, the computed phase spaces (\textit{Full} and \textit{BA}) are comparable in magnitude (see Table~\ref{table:table7} for three EC nuclei: $^{86}$Kr, $^{150}$Sm and $^{207}$Tl).  Consequently, we note only a slight variation, typically of a few factors, in the values of \textit{BA} and \textit{Full} rates other than U1F BD rates. Besides the available phase space values, yet another reason for the enhancement of U1F BD $\lambda_{BA}$ rates is the larger magnitude of total GT strength. Table~\ref{table:table8} shows the computed cumulative GT strength and placement of GT centroid for the six selected nuclei. In this table we depict the GT strength distribution data for the first 10 parent excited states. The cut-off energy in daughter states is 15~MeV. Overall, the larger dissimilarities in available phase spaces between microscopic and \textit{BA} recipes for U1F BD transitions and bigger values of total strength distributions, result in much larger deviation of $\lambda_{BA}$ from $\lambda_{F}$. This later translated to larger magnitudes of R$_{i}$ and $\bar{R}$ values.

\begin{table}[h!]
	\caption{\small Computed phase spaces of \textit{Full} (microscopic) and \textit{BA} rates for allowed and U1F transitions at  selected densities ($\rho \text{Y}_{e}$  [g/cm$^{3}$]) and temperatures (T  [GK]) in stellar environment for three $r$-process BD nuclei.}\label{table:table6} \setlength{\tabcolsep}{5pt}
	\centering {\small
		\resizebox{15cm}{!}{%
			\begin{tabular}{ccccccccccc}
				& & & & & & & \\
				\toprule \multirow{3}{*}{\textbf{T}} &
				\multirow{3}{*}{$\mathbf{\brho Y_{e}}$} & &\multicolumn{2}{c}{\textbf{$^{82}$Ge}}& &
				\multicolumn{2}{c}{\textbf{$^{134}$Te}} & &\multicolumn{2}{c}{\textbf{$^{201}$Re}}\\
				\cmidrule{4-5}  \cmidrule{7-8}  \cmidrule{10-11} & & &
				\multicolumn{1}{c}{\textbf{Full}} &
				\multicolumn{1}{c}{\textbf{BA}} & &
				\multicolumn{1}{c}{\textbf{Full}} &
				\multicolumn{1}{c}{\textbf{BA}} & &
				\multicolumn{1}{c}{\textbf{Full}} &
				\multicolumn{1}{c}{\textbf{BA}} \\
				\cmidrule{3-11}
				\multicolumn{11}{c}{\textbf{U1F}} \\
				\midrule
				1     & 10$^{4}$ &       & 4.30$\times$10$^{+09}$ & 9.60$\times$10$^{+09}$ &       & 1.80$\times$10$^{+09}$ & 2.00$\times$10$^{+09}$ &       & 1.40$\times$10$^{+11}$ & 3.40$\times$10$^{+11}$ \\
				5     & 10$^{4}$ &       & 1.90$\times$10$^{+10}$ & 1.20$\times$10$^{+11}$ &       & 7.60$\times$10$^{+09}$ & 1.60$\times$10$^{+10}$ &       & 1.40$\times$10$^{+11}$ & 3.40$\times$10$^{+11}$ \\
				10    & 10$^{4}$ &       & 1.90$\times$10$^{+10}$ & 1.20$\times$10$^{+11}$ &       & 7.50$\times$10$^{+09}$ & 1.60$\times$10$^{+10}$ &       & 1.40$\times$10$^{+11}$ & 3.40$\times$10$^{+11}$ \\
				20    & 10$^{4}$ &       & 1.80$\times$10$^{+10}$ & 1.20$\times$10$^{+11}$ &       & 7.30$\times$10$^{+09}$ & 1.60$\times$10$^{+10}$ &       & 1.30$\times$10$^{+11}$ & 3.30$\times$10$^{+11}$ \\
				30    & 10$^{4}$ &       & 1.80$\times$10$^{+10}$ & 1.20$\times$10$^{+11}$ &       & 7.00$\times$10$^{+09}$ & 1.50$\times$10$^{+10}$ &       & 1.30$\times$10$^{+11}$ & 3.10$\times$10$^{+11}$ \\
				&       &       &       &       &       &       &       &       &       &  \\
				1     & 10$^{8}$ &       & 4.10$\times$10$^{+09}$ & 9.40$\times$10$^{+09}$ &       & 1.70$\times$10$^{+09}$ & 1.90$\times$10$^{+09}$ &       & 1.40$\times$10$^{+11}$ & 3.30$\times$10$^{+11}$ \\
				5     & 10$^{8}$ &       & 1.80$\times$10$^{+10}$ & 1.20$\times$10$^{+11}$ &       & 7.30$\times$10$^{+09}$ & 1.60$\times$10$^{+10}$ &       & 1.40$\times$10$^{+11}$ & 3.30$\times$10$^{+11}$ \\
				10    & 10$^{8}$ &       & 1.80$\times$10$^{+10}$ & 1.20$\times$10$^{+11}$ &       & 7.40$\times$10$^{+09}$ & 1.60$\times$10$^{+10}$ &       & 1.40$\times$10$^{+11}$ & 3.30$\times$10$^{+11}$ \\
				20    & 10$^{8}$ &       & 1.80$\times$10$^{+10}$ & 1.20$\times$10$^{+11}$ &       & 7.20$\times$10$^{+09}$ & 1.60$\times$10$^{+10}$ &       & 1.30$\times$10$^{+11}$ & 3.20$\times$10$^{+11}$ \\
				30    & 10$^{8}$ &       & 1.70$\times$10$^{+10}$ & 1.20$\times$10$^{+11}$ &       & 7.00$\times$10$^{+09}$ & 1.50$\times$10$^{+10}$ &       & 1.30$\times$10$^{+11}$ & 3.10$\times$10$^{+11}$ \\
				&       &       &       &       &       &       &       &       &       &  \\
				1     & 10$^{11}$ &       & 6.60$\times$10$^{+04}$ & 3.60$\times$10$^{+06}$ &       & 1.60$\times$10$^{-13}$ & 4.10$\times$10$^{-12}$ &       & 1.70$\times$10$^{-37}$ & 2.80$\times$10$^{-42}$ \\
				5     & 10$^{11}$ &       & 3.80$\times$10$^{+07}$ & 9.10$\times$10$^{+09}$ &       & 6.80$\times$10$^{+04}$ & 3.30$\times$10$^{+07}$ &       & 1.90$\times$10$^{-01}$ & 8.10$\times$10$^{-02}$ \\
				10    & 10$^{11}$ &       & 6.30$\times$10$^{+07}$ & 9.70$\times$10$^{+09}$ &       & 1.20$\times$10$^{+06}$ & 5.20$\times$10$^{+07}$ &       & 3.50$\times$10$^{+04}$ & 4.70$\times$10$^{+04}$ \\
				20    & 10$^{11}$ &       & 1.90$\times$10$^{+08}$ & 1.20$\times$10$^{+10}$ &       & 1.90$\times$10$^{+07}$ & 1.60$\times$10$^{+08}$ &       & 4.00$\times$10$^{+07}$ & 8.20$\times$10$^{+07}$ \\
				30    & 10$^{11}$ &       & 4.90$\times$10$^{+08}$ & 1.60$\times$10$^{+10}$ &       & 9.00$\times$10$^{+07}$ & 4.20$\times$10$^{+08}$ &       & 5.90$\times$10$^{+08}$ & 1.30$\times$10$^{+09}$ \\
				\cmidrule{3-11}
				\multicolumn{11}{c}{\textbf{Allowed}} \\
				\cmidrule{3-11}
				1     & 10$^{4}$ &       & 1.80$\times$10$^{+11}$ & 2.70$\times$10$^{+11}$ &       & 2.80$\times$10$^{+10}$ & 3.00$\times$10$^{+10}$ &       & 9.60$\times$10$^{+07}$ & 7.30$\times$10$^{+07}$ \\
				5     & 10$^{4}$ &       & 1.10$\times$10$^{+13}$ & 1.10$\times$10$^{+13}$ &       & 6.90$\times$10$^{+11}$ & 6.90$\times$10$^{+11}$ &       & 9.50$\times$10$^{+07}$ & 7.20$\times$10$^{+07}$ \\
				10    & 10$^{4}$ &       & 1.10$\times$10$^{+13}$ & 1.10$\times$10$^{+13}$ &       & 6.90$\times$10$^{+11}$ & 6.90$\times$10$^{+11}$ &       & 9.10$\times$10$^{+07}$ & 6.90$\times$10$^{+07}$ \\
				20    & 10$^{4}$ &       & 1.10$\times$10$^{+13}$ & 1.10$\times$10$^{+13}$ &       & 6.90$\times$10$^{+11}$ & 6.90$\times$10$^{+11}$ &       & 8.20$\times$10$^{+07}$ & 6.20$\times$10$^{+07}$ \\
				30    & 10$^{4}$ &       & 1.10$\times$10$^{+13}$ & 1.10$\times$10$^{+13}$ &       & 6.80$\times$10$^{+11}$ & 6.80$\times$10$^{+11}$ &       & 7.40$\times$10$^{+07}$ & 5.70$\times$10$^{+07}$ \\
				&       &       &       &       &       &       &       &       &       &  \\
				1     & 10$^{8}$ &       & 1.80$\times$10$^{+11}$ & 2.70$\times$10$^{+11}$ &       & 2.80$\times$10$^{+10}$ & 3.00$\times$10$^{+10}$ &       & 6.70$\times$10$^{+07}$ & 5.10$\times$10$^{+07}$ \\
				5     & 10$^{8}$ &       & 1.10$\times$10$^{+13}$ & 1.10$\times$10$^{+13}$ &       & 6.90$\times$10$^{+11}$ & 6.90$\times$10$^{+11}$ &       & 7.10$\times$10$^{+07}$ & 5.40$\times$10$^{+07}$ \\
				10    & 10$^{8}$ &       & 1.10$\times$10$^{+13}$ & 1.10$\times$10$^{+13}$ &       & 6.90$\times$10$^{+11}$ & 6.90$\times$10$^{+11}$ &       & 7.80$\times$10$^{+07}$ & 5.90$\times$10$^{+07}$ \\
				20    & 10$^{8}$ &       & 1.10$\times$10$^{+13}$ & 1.10$\times$10$^{+13}$ &       & 6.80$\times$10$^{+11}$ & 6.90$\times$10$^{+11}$ &       & 7.80$\times$10$^{+07}$ & 6.00$\times$10$^{+07}$ \\
				30    & 10$^{8}$ &       & 1.10$\times$10$^{+13}$ & 1.10$\times$10$^{+13}$ &       & 6.80$\times$10$^{+11}$ & 6.80$\times$10$^{+11}$ &       & 7.30$\times$10$^{+07}$ & 5.60$\times$10$^{+07}$ \\
				&       &       &       &       &       &       &       &       &       &  \\
				1     & 10$^{11}$ &       & 1.00$\times$10$^{+10}$ & 1.00$\times$10$^{+10}$ &       & 6.50$\times$10$^{+06}$ & 6.50$\times$10$^{+06}$ &       & 3.30$\times$10$^{-68}$ & 5.60$\times$10$^{-70}$ \\
				5     & 10$^{11}$ &       & 4.50$\times$10$^{+12}$ & 4.50$\times$10$^{+12}$ &       & 1.40$\times$10$^{+11}$ & 1.40$\times$10$^{+11}$ &       & 2.60$\times$10$^{-10}$ & 1.20$\times$10$^{-10}$ \\
				10    & 10$^{11}$ &       & 4.60$\times$10$^{+12}$ & 4.60$\times$10$^{+12}$ &       & 1.50$\times$10$^{+11}$ & 1.50$\times$10$^{+11}$ &       & 3.90$\times$10$^{-02}$ & 2.60$\times$10$^{-02}$ \\
				20    & 10$^{11}$ &       & 4.80$\times$10$^{+12}$ & 4.80$\times$10$^{+12}$ &       & 1.60$\times$10$^{+11}$ & 1.60$\times$10$^{+11}$ &       & 1.40$\times$10$^{+03}$ & 1.00$\times$10$^{+03}$ \\
				30    & 10$^{11}$ &       & 5.00$\times$10$^{+12}$ & 5.00$\times$10$^{+12}$ &       & 1.80$\times$10$^{+11}$ & 1.80$\times$10$^{+11}$ &       & 6.20$\times$10$^{+04}$ & 4.60$\times$10$^{+04}$ \\
				\bottomrule
	\end{tabular}}}
\end{table}
\begin{table}[h!]
	\caption{\small Same as Table~\ref{table:table6} but for three $r$-process EC nuclei.}
	\label{table:table7} \setlength{\tabcolsep}{5pt}
	\centering {\small
		\resizebox{15cm}{!}{%
			\begin{tabular}{ccccccccccc}
				& & & & & & & \\
				\toprule \multirow{3}{*}{\textbf{T}} &
				\multirow{3}{*}{$\mathbf{\brho Y_{e}}$} & &\multicolumn{2}{c}{\textbf{$^{86}$Kr}}& &
				\multicolumn{2}{c}{\textbf{$^{150}$Sm}} & &\multicolumn{2}{c}{\textbf{$^{207}$Tl}}\\
				\cmidrule{4-5}  \cmidrule{7-8}  \cmidrule{10-11} & & &
				\multicolumn{1}{c}{\textbf{Full}} &
				\multicolumn{1}{c}{\textbf{BA}} & &
				\multicolumn{1}{c}{\textbf{Full}} &
				\multicolumn{1}{c}{\textbf{BA}} & &
				\multicolumn{1}{c}{\textbf{Full}} &
				\multicolumn{1}{c}{\textbf{BA}} \\
				\cmidrule{3-11}
				\multicolumn{11}{c}{\textbf{U1F}} \\
				\midrule
				1     & 10$^{4}$ &       & 4.60$\times$10$^{+10}$ & 4.60$\times$10$^{+10}$ &       & 5.30$\times$10$^{+01}$ & 1.70$\times$10$^{+10}$ &       & 2.80$\times$10$^{+10}$ & 2.80$\times$10$^{+10}$ \\
				5     & 10$^{4}$ &       & 1.70$\times$10$^{+12}$ & 1.70$\times$10$^{+12}$ &       & 3.60$\times$10$^{+12}$ & 3.80$\times$10$^{+12}$ &       & 2.80$\times$10$^{+10}$ & 2.80$\times$10$^{+10}$ \\
				10    & 10$^{4}$ &       & 1.70$\times$10$^{+12}$ & 1.70$\times$10$^{+12}$ &       & 3.60$\times$10$^{+12}$ & 3.80$\times$10$^{+12}$ &       & 2.80$\times$10$^{+10}$ & 2.80$\times$10$^{+10}$ \\
				20    & 10$^{4}$ &       & 1.70$\times$10$^{+12}$ & 1.70$\times$10$^{+12}$ &       & 3.60$\times$10$^{+12}$ & 3.80$\times$10$^{+12}$ &       & 2.60$\times$10$^{+10}$ & 2.60$\times$10$^{+10}$ \\
				30    & 10$^{4}$ &       & 1.70$\times$10$^{+12}$ & 1.70$\times$10$^{+12}$ &       & 3.60$\times$10$^{+12}$ & 3.80$\times$10$^{+12}$ &       & 2.50$\times$10$^{+10}$ & 2.50$\times$10$^{+10}$ \\
				&       &       &       &       &       &       &       &       &       &  \\
				1     & 10$^{8}$ &       & 4.60$\times$10$^{+10}$ & 4.60$\times$10$^{+10}$ &       & 1.80$\times$10$^{-07}$ & 1.60$\times$10$^{+10}$ &       & 2.70$\times$10$^{+10}$ & 2.70$\times$10$^{+10}$ \\
				5     & 10$^{8}$ &       & 1.70$\times$10$^{+12}$ & 1.70$\times$10$^{+12}$ &       & 3.60$\times$10$^{+12}$ & 3.80$\times$10$^{+12}$ &       & 2.70$\times$10$^{+10}$ & 2.70$\times$10$^{+10}$ \\
				10    & 10$^{8}$ &       & 1.70$\times$10$^{+12}$ & 1.70$\times$10$^{+12}$ &       & 3.60$\times$10$^{+12}$ & 3.80$\times$10$^{+12}$ &       & 2.70$\times$10$^{+10}$ & 2.70$\times$10$^{+10}$ \\
				20    & 10$^{8}$ &       & 1.70$\times$10$^{+12}$ & 1.70$\times$10$^{+12}$ &       & 3.60$\times$10$^{+12}$ & 3.80$\times$10$^{+12}$ &       & 2.60$\times$10$^{+10}$ & 2.60$\times$10$^{+10}$ \\
				30    & 10$^{8}$ &       & 1.70$\times$10$^{+12}$ & 1.70$\times$10$^{+12}$ &       & 3.60$\times$10$^{+12}$ & 3.80$\times$10$^{+12}$ &       & 2.40$\times$10$^{+10}$ & 2.40$\times$10$^{+10}$ \\
				&       &       &       &       &       &       &       &       &       &  \\
				1     & 10$^{11}$ &       & 1.10$\times$10$^{+02}$ & 1.10$\times$10$^{+02}$ &       & 8.20$\times$10$^{-16}$ & 3.80$\times$10$^{-13}$ &       & 2.60$\times$10$^{-56}$ & 2.60$\times$10$^{-56}$ \\
				5     & 10$^{11}$ &       & 5.90$\times$10$^{+11}$ & 5.90$\times$10$^{+11}$ &       & 9.60$\times$10$^{+11}$ & 9.60$\times$10$^{+11}$ &       & 2.10$\times$10$^{-05}$ & 2.10$\times$10$^{-05}$ \\
				10    & 10$^{11}$ &       & 5.90$\times$10$^{+11}$ & 5.90$\times$10$^{+11}$ &       & 9.80$\times$10$^{+11}$ & 9.80$\times$10$^{+11}$ &       & 2.40$\times$10$^{+02}$ & 2.40$\times$10$^{+02}$ \\
				20    & 10$^{11}$ &       & 6.30$\times$10$^{+11}$ & 6.30$\times$10$^{+11}$ &       & 1.10$\times$10$^{+12}$ & 1.10$\times$10$^{+12}$ &       & 1.90$\times$10$^{+06}$ & 1.90$\times$10$^{+06}$ \\
				30    & 10$^{11}$ &       & 6.80$\times$10$^{+11}$ & 6.80$\times$10$^{+11}$ &       & 1.20$\times$10$^{+12}$ & 1.20$\times$10$^{+12}$ &       & 4.90$\times$10$^{+07}$ & 4.90$\times$10$^{+07}$ \\
				\cmidrule{3-11}
				\multicolumn{11}{c}{\textbf{Allowed}} \\
				\cmidrule{3-11}
				1     & 10$^{4}$ &       & 9.40$\times$10$^{+09}$ & 9.50$\times$10$^{+09}$ &       & 1.80$\times$10$^{+11}$ & 1.80$\times$10$^{+11}$ &       & 5.80$\times$10$^{+09}$ & 5.90$\times$10$^{+09}$ \\
				5     & 10$^{4}$ &       & 4.70$\times$10$^{+11}$ & 4.70$\times$10$^{+11}$ &       & 4.00$\times$10$^{+12}$ & 4.00$\times$10$^{+12}$ &       & 5.70$\times$10$^{+09}$ & 5.80$\times$10$^{+09}$ \\
				10    & 10$^{4}$ &       & 4.70$\times$10$^{+11}$ & 4.70$\times$10$^{+11}$ &       & 4.00$\times$10$^{+12}$ & 4.00$\times$10$^{+12}$ &       & 5.60$\times$10$^{+09}$ & 5.70$\times$10$^{+09}$ \\
				20    & 10$^{4}$ &       & 4.70$\times$10$^{+11}$ & 4.70$\times$10$^{+11}$ &       & 4.00$\times$10$^{+12}$ & 4.00$\times$10$^{+12}$ &       & 5.20$\times$10$^{+09}$ & 5.30$\times$10$^{+09}$ \\
				30    & 10$^{4}$ &       & 4.60$\times$10$^{+11}$ & 4.60$\times$10$^{+11}$ &       & 4.00$\times$10$^{+12}$ & 4.00$\times$10$^{+12}$ &       & 4.80$\times$10$^{+09}$ & 4.90$\times$10$^{+09}$ \\
				&       &       &       &       &       &       &       &       &       &  \\
				1     & 10$^{8}$ &       & 9.30$\times$10$^{+09}$ & 9.30$\times$10$^{+09}$ &       & 1.80$\times$10$^{+11}$ & 1.80$\times$10$^{+11}$ &       & 4.90$\times$10$^{+09}$ & 5.00$\times$10$^{+09}$ \\
				5     & 10$^{8}$ &       & 4.70$\times$10$^{+11}$ & 4.70$\times$10$^{+11}$ &       & 4.00$\times$10$^{+12}$ & 4.00$\times$10$^{+12}$ &       & 5.00$\times$10$^{+09}$ & 5.10$\times$10$^{+09}$ \\
				10    & 10$^{8}$ &       & 4.70$\times$10$^{+11}$ & 4.70$\times$10$^{+11}$ &       & 4.00$\times$10$^{+12}$ & 4.00$\times$10$^{+12}$ &       & 5.20$\times$10$^{+09}$ & 5.20$\times$10$^{+09}$ \\
				20    & 10$^{8}$ &       & 4.70$\times$10$^{+11}$ & 4.70$\times$10$^{+11}$ &       & 4.00$\times$10$^{+12}$ & 4.00$\times$10$^{+12}$ &       & 5.10$\times$10$^{+09}$ & 5.10$\times$10$^{+09}$ \\
				30    & 10$^{8}$ &       & 4.60$\times$10$^{+11}$ & 4.60$\times$10$^{+11}$ &       & 4.00$\times$10$^{+12}$ & 4.00$\times$10$^{+12}$ &       & 4.80$\times$10$^{+09}$ & 4.80$\times$10$^{+09}$ \\
				&       &       &       &       &       &       &       &       &       &  \\
				1     & 10$^{11}$ &       & 8.00$\times$10$^{+00}$ & 8.00$\times$10$^{+00}$ &       & 1.20$\times$10$^{+08}$ & 1.20$\times$10$^{+08}$ &       & 1.40$\times$10$^{-57}$ & 1.40$\times$10$^{-57}$ \\
				5     & 10$^{11}$ &       & 1.20$\times$10$^{+11}$ & 1.20$\times$10$^{+11}$ &       & 1.40$\times$10$^{+12}$ & 1.40$\times$10$^{+12}$ &       & 1.50$\times$10$^{-06}$ & 1.50$\times$10$^{-06}$ \\
				10    & 10$^{11}$ &       & 1.20$\times$10$^{+11}$ & 1.20$\times$10$^{+11}$ &       & 1.40$\times$10$^{+12}$ & 1.40$\times$10$^{+12}$ &       & 2.20$\times$10$^{+01}$ & 2.20$\times$10$^{+01}$ \\
				20    & 10$^{11}$ &       & 1.30$\times$10$^{+11}$ & 1.30$\times$10$^{+11}$ &       & 1.50$\times$10$^{+12}$ & 1.50$\times$10$^{+12}$ &       & 2.30$\times$10$^{+05}$ & 2.30$\times$10$^{+05}$ \\
				30    & 10$^{11}$ &       & 1.50$\times$10$^{+11}$ & 1.50$\times$10$^{+11}$ &       & 1.60$\times$10$^{+12}$ & 1.60$\times$10$^{+12}$ &       & 6.80$\times$10$^{+06}$ & 6.90$\times$10$^{+06}$ \\
				\bottomrule
	\end{tabular}}}
\end{table}
\begin{table}[h!]
	\centering \caption{\small Computed total GT strength ($\Sigma B$ in arbitrary units) and centroid ($\bar{E}$ in MeV units) values for selected nuclei in EC (left panel) and BD (right panel) directions of U1F and allowed GT transitions for 10 parent excited states. The energy cutoff in daughter states is 15 MeV.}\label{table:table8}
	\setlength{\aboverulesep}{0pt} \setlength{\belowrulesep}{0pt} \setlength{\tabcolsep}{8pt} \setlength{\arrayrulewidth}{3pt}
	\renewcommand{\arraystretch}{1.0} 
	{\small
		\centering  \hspace{20pt}
		\resizebox{1\textwidth}{!}{%
			\begin{tabular}{ccccc|ccccc}
				\toprule
				\multicolumn{5}{|c}{ \textbf{EC} } & \multicolumn{5}{c|}{ \textbf{BD} }\\[0.4ex]
				\midrule[2pt]
				\midrule[1pt] \\[-1.0em]
				\multicolumn{5}{c}{\textbf{$^{86}$Kr}} & \multicolumn{5}{c}{\textbf{$^{82}$Ge}} \\[0.4ex]
				\midrule[1pt] \\[-1.0em]
				\multicolumn{1}{c}{$\Sigma B(U1F)_{+}$} &
				\multicolumn{1}{c}{$\Sigma B(GT)_{+}$} & &
				\multicolumn{1}{c}{$\bar{E}_{+} [U1F]$} &
				\multicolumn{1}{c}{$\bar{E}_{+} [GT]$} & 
				\multicolumn{1}{c}{$\Sigma B(U1F)_{-}$} &
				\multicolumn{1}{c}{$\Sigma B(GT)_{-}$} & &
				\multicolumn{1}{c}{$\bar{E}_{-} [U1F]$} &
				\multicolumn{1}{c}{$\bar{E}_{-} [GT]$} \\
				\cmidrule{1-2}  \cmidrule{4-5} \cmidrule{6-7} \cmidrule{9-10}
				13.95 & 9.57  &       & 3.31  & 2.24  & 52.30 & 49.53 &       & 9.09  & 11.64 \\
				41.83 & 29.66 &       & 6.02  & 5.11  & 88.73 & 49.58 &       & 24.31 & 6.00 \\
				42.28 & 28.33 &       & 10.00 & 6.24  & 158.59 & 51.28 &       & 12.60 & 6.42 \\
				40.74 & 34.48 &       & 7.16  & 4.50  & 114.55 & 49.49 &       & 11.20 & 4.08 \\
				42.99 & 47.55 &       & 11.32 & 5.20  & 184.16 & 96.27 &       & 11.17 & 4.67 \\
				50.49 & 48.33 &       & 10.81 & 3.77  & 551.94 & 96.41 &       & 11.39 & 6.63 \\
				52.75 & 51.71 &       & 12.79 & 9.41  & 315.25 & 126.68 &       & 9.40  & 8.50 \\
				38.32 & 38.16 &       & 12.19 & 9.56  & 173.40 & 165.13 &       & 12.89 & 6.07 \\
				57.27 & 31.24 &       & 11.35 & 10.51 & 227.86 & 135.00 &       & 11.52 & 3.05 \\
				54.70 & 48.41 &       & 12.76 & 7.86  & 469.73 & 148.15 &       & 10.87 & 3.49 \\
				\midrule[1pt]
				\midrule[1pt] \\[-1.0em]
				\multicolumn{5}{c}{\textbf{$^{150}$Sm}} & \multicolumn{5}{c}{\textbf{$^{134}$Te}} \\[0.4ex]
				\midrule[1pt] \\[-1.0em]
				\multicolumn{1}{c}{$\Sigma B(U1F)_{+}$} &
				\multicolumn{1}{c}{$\Sigma B(GT)_{+}$} & &
				\multicolumn{1}{c}{$\bar{E}_{+} [U1F]$} &
				\multicolumn{1}{c}{$\bar{E}_{+} [GT]$} & 
				\multicolumn{1}{c}{$\Sigma B(U1F)_{-}$} &
				\multicolumn{1}{c}{$\Sigma B(GT)_{-}$} & &
				\multicolumn{1}{c}{$\bar{E}_{-} [U1F]$} &
				\multicolumn{1}{c}{$\bar{E}_{-} [GT]$} \\
				\cmidrule{1-2}  \cmidrule{4-5} \cmidrule{6-7} \cmidrule{9-10}
				30.33 & 17.46 &       & 8.56  & 10.86 & 249.29 & 81.30 &       & 6.81  & 8.95 \\
				35.89 & 36.49 &       & 8.63  & 3.66  & 361.76 & 318.58 &       & 8.78  & 12.47 \\
				37.43 & 56.22 &       & 9.03  & 6.31  & 997.63 & 582.70 &       & 8.81  & 11.31 \\
				50.19 & 62.18 &       & 9.14  & 5.08  & 1060.19 & 480.44 &       & 8.98  & 9.08 \\
				47.50 & 41.00 &       & 9.49  & 5.50  & 1299.62 & 245.19 &       & 13.07 & 8.81 \\
				52.89 & 67.49 &       & 10.07 & 6.22  & 3441.34 & 323.53 &       & 10.08 & 9.84 \\
				54.93 & 62.45 &       & 9.58  & 6.97  & 1223.27 & 514.90 &       & 10.70 & 10.01 \\
				55.25 & 55.47 &       & 10.84 & 7.32  & 1569.68 & 1621.17 &       & 9.45  & 10.29 \\
				67.34 & 60.42 &       & 10.68 & 7.61  & 1829.43 & 1807.52 &       & 9.40  & 10.52 \\
				63.47 & 88.16 &       & 11.21 & 8.57  & 3165.09 & 469.26 &       & 10.14 & 10.28 \\
				\midrule[1pt]
				\midrule[1pt] \\[-1.0em]
				\multicolumn{5}{c}{\textbf{$^{207}$Tl}} & \multicolumn{5}{c}{\textbf{$^{201}$Re}} \\[0.4ex]
				\midrule[1pt] \\[-1.0em]
				\multicolumn{1}{c}{$\Sigma B(U1F)_{+}$} &
				\multicolumn{1}{c}{$\Sigma B(GT)_{+}$} & &
				\multicolumn{1}{c}{$\bar{E}_{+} [U1F]$} &
				\multicolumn{1}{c}{$\bar{E}_{+} [GT]$} & 
				\multicolumn{1}{c}{$\Sigma B(U1F)_{-}$} &
				\multicolumn{1}{c}{$\Sigma B(GT)_{-}$} & &
				\multicolumn{1}{c}{$\bar{E}_{-} [U1F]$} &
				\multicolumn{1}{c}{$\bar{E}_{-} [GT]$} \\
				\cmidrule{1-2}  \cmidrule{4-5} \cmidrule{6-7} \cmidrule{9-10}
				29.91 & 28.72 &       & 13.02 & 12.60 & 171.66 & 138.16 &       & 11.87 & 8.57 \\
				39.78 & 27.45 &       & 12.92 & 13.23 & 205.79 & 142.55 &       & 10.09 & 8.73 \\
				46.08 & 32.14 &       & 13.83 & 14.20 & 207.82 & 137.39 &       & 10.12 & 11.69 \\
				3.53  & 40.29 &       & 9.31  & 11.75 & 279.69 & 138.29 &       & 9.34  & 10.84 \\
				6.69  & 43.60 &       & 9.93  & 9.34  & 172.59 & 138.23 &       & 9.26  & 12.90 \\
				4.80  & 50.39 &       & 9.81  & 13.17 & 171.55 & 138.15 &       & 10.91 & 13.04 \\
				7.19  & 51.26 &       & 10.34 & 11.63 & 172.55 & 138.29 &       & 12.31 & 11.75 \\
				5.60  & 53.59 &       & 9.70  & 12.95 & 172.66 & 138.13 &       & 12.55 & 13.17 \\
				11.22 & 46.21 &       & 9.76  & 12.40 & 172.50 & 137.27 &       & 12.53 & 12.14 \\
				3.55  & 51.44 &       & 10.07 & 12.69 & 208.95 & 138.13 &       & 11.13 & 13.24 \\
				\bottomrule
	\end{tabular}}}
\end{table}

For the BD rates (U1F and allowed) of N = 50 and N = 82 nuclei (Table~\ref{table:table2}),  the calculated  value of R$_{i}$ equals 1.00 at T = 1 [GK]. This implies that at this temperature, both $\lambda_{F}$ and $\lambda_{BA}$ are identical. This core temperature corresponds roughly to the neon burning phases of the star. Consequently, it is concluded that \textit{BA} hypothesis may be safely applied to stellar BD rates till neon burning phases of massive stars. For heavy nuclei (e.g., $^{201}$Re) the \textit{BA} fails even at T = 1 [GK]. Table~\ref{table:table3} computes much smaller values of $\bar{R}$ for EC rates. This means that EC rates are less affected by usage of \textit{BA} hypothesis as compared to BD rates. 

Figs.~(\ref{Fig. 3}$-$\ref{Fig. 8})  show that 
BD and EC rates increase as the core temperature rises due to rise in the occupation probability of parent excited states. Consequently, contribution of the  partial rates to the total weak rates becomes significant. 
The magnitude of BD rates decreases with rise in density due to a decrease in available phase space. 
EC rates enhance when electron chemical potential goes higher with the increase in density.

\section{Conclusions}
\label{sec:conclusions}
The effectiveness of application of the \textit{BA} hypothesis for calculation of allowed weak rates of  heavy nuclei, in general, and forbidden rates, in particular,   was missing in literature. In the present work, the impact of applying the \textit{BA} hypothesis on calculation of stellar rates was investigated. The chosen range of nuclei, having A = 70 -- 208 and Z = 27 -- 82, has vast applications in the  
$r$-process nucleosynthesis.  The pn-QRPA model was employed to 
evaluate \textit{Full} and \textit{BA} rates over a wide range of temperature (1$-$30) [GK]  and density (10$-$10$^{11}$)  [g/cm$^{3}$] for allowed GT and U1F rates. 
 The comparison of 
microscopic state-to-state calculated rates with those obtained by applying the \textit{BA} hypothesis, indicates a sizeable change 
in both rate values, particularly, in the case of U1F BD rates. Here the \textit{BA} hypothesis is found to have the strongest 
effect with deviations exceeding four to five orders of magnitude. While, in the case of other rates, 
including BD and EC rates of allowed transitions and EC rates of U1F transitions, the deviation was rather small. According to our investigation, the total \textit{BA}  rates (including both allowed and U1F contributions) deviate from total \textit{Full}  rates by  an order of magnitude or more. This order of magnitude deviation was recorded both for EC and BD rates. Core-collapse simulators might find this information useful for modeling purposes.  Our findings indicate that the \textit{BA} hypothesis may be safely applied for BD rates, until the core temperature reaches  T = 1 [GK] for all density regions. The weak rates based on the \textit{BA} hypothesis, in general, start to deviate from the microscopically calculated rates when core densities and temperatures exceed $10^{4}$ 
[g/cm$^{3}$] and 5 [GK], respectively.

Noticeable mentions of the current investigation include:

$\odot$ For BD rates, more than 1 (5) orders of magnitude difference in allowed (U1F) rates is noted by usage of \textit{BA}. Forbidden transitions are more affected by usage of \textit{BA} hypothesis when compared with the allowed GT transitions. 

$\odot$ For small BD rates ($\lambda$ $<$ 10$^{-5}$ [s$^{-1}$]), at high core temperatures (T $>$ 10 [GK]) and densities ($\rho$Y$_{e} >$ 10$^{8}$ [g/cm$^{3}$]), more than 5 orders of magnitude difference was reported between \textit{Full} and \textit{BA} rates. These temperature-density conditions correspond roughly to the silicon burning phase of the star~\cite{Pran11}.

$\odot$ For EC rates, more than 2 (1) orders of magnitude difference in allowed (U1F) rates is noted by usage of \textit{BA} hypothesis.  

$\odot$ For small EC rates ($\lambda$ $<$ 10$^{-5}$ [s$^{-1}$]), at low core temperatures (T $\le$ 1 [GK]) and densities ($\rho$Y$_{e} \le $ 10$^{4}$ [g/cm$^{3}$]), more than 20 orders of magnitude difference was reported between \textit{Full} and \textit{BA} rates. These physical conditions correspond roughly to C-burning and pre-C-burning phase of the star~\cite{Pran11}.

$\odot$ The total \textit{BA} rates, including both allowed and U1F contributions, deviate by an order of magnitude or more from the total \textit{Full} rates.

From this result, it is evident that the \textit{BA} hypothesis has a significant impact on the accuracy and reliability of nuclear physics inputs used for the simulations of $r$-process nucleosynthesis, where, contributions of forbidden transitions to total rates become large. The current investigation merits due consideration prior to the application of  \textit{BA} hypothesis for the weak rate calculations.  In the Appendix we show ratios and average deviations of 9 new BD and EC nuclei. Detailed data of these and remaining nuclei may be requested from the corresponding author. 

\ack J.-U. Nabi would like to acknowledge the support of the Higher Education Commission Pakistan through 
Project \# 20-15394/NRPU/R\&D/HEC/2021.
\clearpage
\appendix
\section*{Appendix. Ratios and averaged ratios between \textit{Full} and \textit{BA} rates}

Comparison of \textit{BA} ($\lambda_{BA}$) and \textit{Full} ($\lambda_{F}$) BD rates for three selected nuclei as a function of core temperature (T [GK]) and density ($\rho \text{Y}_{e}$  [g/cm$^{3}$]). Ratios (R$_{i}$) and average deviation ($\bar{R}$) are given separately for allowed (All) and forbidden (U1F) transitions and are defined in Eq.~\ref{Ri} and Eq.~\ref{Rbar}, respectively. Computed R$_{i}$ values, where $\lambda_{F}$ and/or $\lambda_{BA}$ rates are less than 10$^{-5}$ s$^{-1}$, are marked with  *. $\bar{R}^{(*)}$ are computed average deviations excluding ratios marked with *.
\setcounter{section}{1} 

\begin{table}[h]
	\caption{\small R$_{i}$ [All] and R$_{i}$ [U1F] for three BD nuclei. See text for explanation of symbols.
	}\label{table:table9} \setlength{\tabcolsep}{2pt} 
	\centering {\small
		\resizebox{17cm}{!}{%
			\begin{tabular}{cc|c|c|c|c|c|c|c|c|c|c|c|c}
				\toprule \multirow{3}{*}{\textbf{T}} &
				\multirow{3}{*}{$\mathbf{\brho Y_{e}}$} & \multicolumn{4}{c|}{\textbf{$^{79}$Zn}}& \multicolumn{4}{c|}{\textbf{$^{80}$Zn}} & \multicolumn{4}{c}{\textbf{$^{96}$Zr}}\\
				\cmidrule{3-6}  \cmidrule{7-10}  \cmidrule{11-14} & & 
				\multicolumn{2}{c|}{\bf{R$_{i}$} [All]} &
				\multicolumn{2}{c|}{\bf{R$_{i}$} [U1F]} & 
				\multicolumn{2}{c|}{\bf{R$_{i}$} [All]} &
				\multicolumn{2}{c|}{\bf{R$_{i}$} [U1F]} & 
				\multicolumn{2}{c|}{\bf{R$_{i}$} [All]} &
				\multicolumn{2}{c}{\bf{R$_{i}$} [U1F]} \\ & & 
				\multicolumn{1}{c|}{\bf{$\lambda_{F} \ge \lambda_{BA}$}} &
				\multicolumn{1}{c|}{\bf{$\lambda_{BA}>\lambda_{F}$}} &
				\multicolumn{1}{c|}{\bf{$\lambda_{F} \ge \lambda_{BA}$}} &
				\multicolumn{1}{c|}{\bf{$\lambda_{BA}>\lambda_{F}$}}  & 
				\multicolumn{1}{c|}{\bf{$\lambda_{F} \ge \lambda_{BA}$}} &
				\multicolumn{1}{c|}{\bf{$\lambda_{BA}>\lambda_{F}$}} & 
				\multicolumn{1}{c|}{\bf{$\lambda_{F} \ge \lambda_{BA}$}} &
				\multicolumn{1}{c|}{\bf{$\lambda_{BA}>\lambda_{F}$}} &
				\multicolumn{1}{c|}{\bf{$\lambda_{F} \ge \lambda_{BA}$}}  & 
				\multicolumn{1}{c|}{\bf{$\lambda_{BA}>\lambda_{F}$}} &
				\multicolumn{1}{c|}{\bf{$\lambda_{F} \ge \lambda_{BA}$}}  & 
				\multicolumn{1}{c}{\bf{$\lambda_{BA}>\lambda_{F}$}} \\
				\hline
				1     & 10$^{4}$ & 1.0  &       &       & 1.0  &       & 1.0  &       & 1.0  & 1.0$^{(*)}$ &     &       & 1.0$^{(*)}$ \\
				10    & 10$^{4}$ & 1.9  &       &       & 21.6 &       & 1.4  &       & 11.7 & 3.7  &   &       & 11272.0 \\
				30    & 10$^{4}$ & 6.6  &       &       & 345.9 &       & 3.2  &       & 373.3 &    & 6.8  &       & 158489.3 \\
				&       &       &       &       &       &       &       &       &       &       &       &       &  \\
				1     & 10$^{8}$ & 1.0  &       &       & 6.6  &       & 1.0  &       & 10.0 & 2.5$^{(*)}$  &          &       & 2.4$^{(*)}$ \\
				10    & 10$^{8}$ & 2.1  &       &       & 77.8 &       & 1.5  &       & 32.3 & 3.3  &    &       & 23173.9 \\
				30    & 10$^{8}$ & 6.6  &       &       & 701.5 &       & 3.2  &       & 379.3 &   & 6.9  &       & 130017.0 \\
				&       &       &       &       &       &       &       &       &       &       &       &       &  \\
				1     & 10$^{11}$ &--- & --- & --- &--- & --- & --- & --- & ---&--- & --- &--- & --- \\
				10    & 10$^{11}$ & 6.8$^{(*)}$  &       &       & 15995.6$^{(*)}$ &       & 5.0$^{(*)}$  &       & 35892.2$^{(*)}$ &    & 30.1$^{(*)}$ &       & 70957.8$^{(*)}$ \\
				30    & 10$^{11}$ & 6.3  &       &       & 31260.8 &       & 23.0 &       & 67920.4 &   & 95.1 &       & 65917.4$^{(*)}$ \\
				\hline
				\multicolumn{2}{c|}{$\mathbf{\bar{R}}$} &   \multicolumn{2}{c|}{4.1} & \multicolumn{2}{c|}{6051.4} & \multicolumn{2}{c|}{4.9} & \multicolumn{2}{c|}{13077.5} & \multicolumn{2}{c|}{18.7} & \multicolumn{2}{c}{57478.9} \\
				\hline
				\multicolumn{2}{c|}{$\mathbf{\bar{R}^{(*)}}$} &   \multicolumn{2}{c|}{3.7} & \multicolumn{2}{c|}{4630.8} & \multicolumn{2}{c|}{4.9} & \multicolumn{2}{c|}{9818.3} & \multicolumn{2}{c|}{23.2} & \multicolumn{2}{c}{80738.0} 
	\end{tabular}}}
\end{table}
\begin{table}[h]
	\caption{\small R$_{i}$ [All] and R$_{i}$ [U1F] for three BD nuclei. See text for explanation of symbols.
	}\label{table:table10} \setlength{\tabcolsep}{2pt} 
	\centering {\small
		\resizebox{17cm}{!}{%
			\begin{tabular}{cc|c|c|c|c|c|c|c|c|c|c|c|c}
				\toprule \multirow{3}{*}{\textbf{T}} &
				\multirow{3}{*}{$\mathbf{\brho Y_{e}}$} &\multicolumn{4}{c|}{\textbf{$^{100}$Mo}}& 
				\multicolumn{4}{c|}{\textbf{$^{124}$Sn}}  &\multicolumn{4}{c}{\textbf{$^{130}$Te}}\\
				\cmidrule{3-6}  \cmidrule{7-10}  \cmidrule{11-14} & & 
				\multicolumn{2}{c|}{\bf{R$_{i}$} [All]} &
				\multicolumn{2}{c|}{\bf{R$_{i}$} [U1F]} & 
				\multicolumn{2}{c|}{\bf{R$_{i}$} [All]} &
				\multicolumn{2}{c|}{\bf{R$_{i}$} [U1F]} & 
				\multicolumn{2}{c|}{\bf{R$_{i}$} [All]} &
				\multicolumn{2}{c}{\bf{R$_{i}$} [U1F]} \\ & & 
				\multicolumn{1}{c|}{\bf{$\lambda_{F} \ge \lambda_{BA}$}} &
				\multicolumn{1}{c|}{\bf{$\lambda_{BA}>\lambda_{F}$}} &
				\multicolumn{1}{c|}{\bf{$\lambda_{F} \ge \lambda_{BA}$}} &
				\multicolumn{1}{c|}{\bf{$\lambda_{BA}>\lambda_{F}$}}  & 
				\multicolumn{1}{c|}{\bf{$\lambda_{F} \ge \lambda_{BA}$}} &
				\multicolumn{1}{c|}{\bf{$\lambda_{BA}>\lambda_{F}$}} & 
				\multicolumn{1}{c|}{\bf{$\lambda_{F} \ge \lambda_{BA}$}} &
				\multicolumn{1}{c|}{\bf{$\lambda_{BA}>\lambda_{F}$}} &
				\multicolumn{1}{c|}{\bf{$\lambda_{F} \ge \lambda_{BA}$}}  & 
				\multicolumn{1}{c|}{\bf{$\lambda_{BA}>\lambda_{F}$}} &
				\multicolumn{1}{c|}{\bf{$\lambda_{F} \ge \lambda_{BA}$}}  & 
				\multicolumn{1}{c}{\bf{$\lambda_{BA}>\lambda_{F}$}} \\
				\hline
				1     & 10$^{4}$ & 4.7$^{(*)}$  &       &       & 1.0$^{(*)}$  & 1.0  &       &       & 1.0  & 146.6$^{(*)}$ &       & 12.3$^{(*)}$ &     \\
				10    & 10$^{4}$ & 3.6  &       &       & 2041.7$^{(*)}$ & 38.2 &       &       & 76.2 & 41.2 &       &       & 25.5 \\
				30    & 10$^{4}$ &       & 7.8  &       & 56623.9 & 11.4 &       &       & 1419.1 & 16.6 &       &       & 8953.6 \\
				&       &       &       &       &       &       &       &       &       &       &       &       &  \\
				1     & 10$^{8}$ & 5.7$^{(*)}$  &       &       & 33.3$^{(*)}$ & 1.0  &       &       & 1.0  & 263.6$^{(*)}$ &       & 8.2$^{(*)}$  &     \\
				10    & 10$^{8}$ & 3.3  &       &       & 1798.9$^{(*)}$ & 40.0 &       &       & 1205.0 & 38.5 &       &       & 28119.0$^{(*)}$ \\
				30    & 10$^{8}$ &       & 7.9  &       & 58479.0 & 11.4 &       &       & 9311.1 & 16.6 &       &       & 85113.8 \\
				&       &       &       &       &       &       &       &       &       &       &       &       &  \\
				1     & 10$^{11}$ & ---     & ---     & ---     & ---     & ---     & ---     & ---     & ---     & ---     & ---     & ---     & --- \\
				10    & 10$^{11}$ &       & 28.3$^{(*)}$ &       & 14060.5$^{(*)}$ & 14.7$^{(*)}$ &       &       & 47643.1$^{(*)}$ & 20.4$^{(*)}$ &       &       & 54075.4$^{(*)}$ \\
				30    & 10$^{11}$ &       & 98.9 &       & 240990.5$^{(*)}$ & 10.9 &       &       & 36728.2 & 13.6 &       &       & 67764.2 \\
				\hline
				\multicolumn{2}{c|}{$\mathbf{\bar{R}}$} & \multicolumn{2}{c|}{20.0} & \multicolumn{2}{c|}{46753.6} & \multicolumn{2}{c|}{16.1} & \multicolumn{2}{c|}{12048.1} & \multicolumn{2}{c|}{69.6} & \multicolumn{2}{c}{30509.0} \\
				\hline
				\multicolumn{2}{c|}{$\mathbf{\bar{R}^{(*)}}$} &     \multicolumn{2}{c|}{24.3} & \multicolumn{2}{c|}{57551.5} & \multicolumn{2}{c|}{16.3} & \multicolumn{2}{c|}{6963.1} & \multicolumn{2}{c|}{25.3} & \multicolumn{2}{c}{40464.3} 
	\end{tabular}}}
\end{table}
\begin{table}[h]
	\caption{\small R$_{i}$ [All] and R$_{i}$ [U1F] for three BD nuclei. See text for explanation of symbols.
	}\label{table:table11} \setlength{\tabcolsep}{2pt} 
	\centering {\small
		\resizebox{17cm}{!}{%
			\begin{tabular}{cc|c|c|c|c|c|c|c|c|c|c|c|c}
				\toprule \multirow{3}{*}{\textbf{T}} &
				\multirow{3}{*}{$\mathbf{\brho Y_{e}}$} &\multicolumn{4}{c|}{\textbf{$^{136}$Xe}}& 
				\multicolumn{4}{c|}{\textbf{$^{150}$Nd}}  &\multicolumn{4}{c}{\textbf{$^{202}$Os}}\\
				\cmidrule{3-6}  \cmidrule{7-10}  \cmidrule{11-14} & & 
				\multicolumn{2}{c|}{\bf{R$_{i}$} [All]} &
				\multicolumn{2}{c|}{\bf{R$_{i}$} [U1F]} & 
				\multicolumn{2}{c|}{\bf{R$_{i}$} [All]} &
				\multicolumn{2}{c|}{\bf{R$_{i}$} [U1F]} & 
				\multicolumn{2}{c|}{\bf{R$_{i}$} [All]} &
				\multicolumn{2}{c}{\bf{R$_{i}$} [U1F]} \\ & & 
				\multicolumn{1}{c|}{\bf{$\lambda_{F} \ge \lambda_{BA}$}} &
				\multicolumn{1}{c|}{\bf{$\lambda_{BA}>\lambda_{F}$}} &
				\multicolumn{1}{c|}{\bf{$\lambda_{F} \ge \lambda_{BA}$}} &
				\multicolumn{1}{c|}{\bf{$\lambda_{BA}>\lambda_{F}$}}  & 
				\multicolumn{1}{c|}{\bf{$\lambda_{F} \ge \lambda_{BA}$}} &
				\multicolumn{1}{c|}{\bf{$\lambda_{BA}>\lambda_{F}$}} & 
				\multicolumn{1}{c|}{\bf{$\lambda_{F} \ge \lambda_{BA}$}} &
				\multicolumn{1}{c|}{\bf{$\lambda_{BA}>\lambda_{F}$}} &
				\multicolumn{1}{c|}{\bf{$\lambda_{F} \ge \lambda_{BA}$}}  & 
				\multicolumn{1}{c|}{\bf{$\lambda_{BA}>\lambda_{F}$}} &
				\multicolumn{1}{c|}{\bf{$\lambda_{F} \ge \lambda_{BA}$}}  & 
				\multicolumn{1}{c}{\bf{$\lambda_{BA}>\lambda_{F}$}} \\
				\hline
				1     & 10$^{4}$ & 133.0$^{(*)}$ &       &       & 7.2$^{(*)}$  & 9.0$^{(*)}$  &       &       & 1.0$^{(*)}$  & 1.0  &       &       & 1.0 \\
				10    & 10$^{4}$ & 46.7 &       &       & 13091.8 & 2.4  &       &       & 251.2$^{(*)}$ & 21.9 &       &       & 1472.3 \\
				30    & 10$^{4}$ & 28.7 &       &       & 130017.0 &       & 1.6  &       & 20844.9 & 19.3 &       &       & 2454.7 \\
				&       &       &       &       &       &       &       &       &       &       &       &       &  \\
				1     & 10$^{8}$ & 150.0$^{(*)}$ &       &       & 1.2$^{(*)}$  & 11.5$^{(*)}$ &       &       & 1.0$^{(*)}$  & 1.0$^{(*)}$  &       &       & 1.0$^{(*)}$ \\
				10    & 10$^{8}$ & 44.3 &       &       & 8260.4 & 2.0  &       &       & 7620.8$^{(*)}$ & 22.3 &       &       & 26977.40 \\
				30    & 10$^{8}$ & 28.6 &       &       & 314050.9 &  & 1.7  &       & 135831.3 & 19.3 &       &       & 62517.3 \\
				&       &       &       &       &       &       &       &       &       &       &       &       &  \\
				1     & 10$^{11}$ & ---     & ---     & ---     & ---     & ---     & ---     & ---     & ---     & ---     & ---     & ---     & --- \\
				10    & 10$^{11}$ & 29.9$^{(*)}$ &       &       & 68706.8$^{(*)}$ &   & 3.2 $^{(*)}$ &       & 33728.7$^{(*)}$ & 19.1$^{(*)}$ &       &       & 16943.4$^{(*)}$ \\
				30    & 10$^{11}$ & 26.5 &       &       & 135207.3$^{(*)}$ &  & 2.8  &       & 113501.1$^{(*)}$ & 19.1 &       &       & 344349.9$^{(*)}$ \\
				\hline
				\multicolumn{2}{c|}{$\mathbf{\bar{R}}$} & \multicolumn{2}{c|}{61.0} & \multicolumn{2}{c|}{83667.8} & \multicolumn{2}{c|}{4.3} & \multicolumn{2}{c|}{38972.5} & \multicolumn{2}{c|}{15.4} & \multicolumn{2}{c}{56839.6} \\
				\hline
				\multicolumn{2}{c|}{$\mathbf{\bar{R}^{(*)}}$} &    \multicolumn{2}{c|}{35.0} & \multicolumn{2}{c|}{116355.0} & \multicolumn{2}{c|}{2.1} & \multicolumn{2}{c|}{78338.1} & \multicolumn{2}{c|}{17.1} & \multicolumn{2}{c}{18684.5} 
	\end{tabular}}}
\end{table}
\begin{table}[h]
	\caption{\small R$_{i}$ [All] and R$_{i}$ [U1F] for three EC nuclei. See text for explanation of symbols.
	}\label{table:table12} \setlength{\tabcolsep}{2pt} 
	\centering {\small
		\resizebox{17cm}{!}{%
			\begin{tabular}{cc|c|c|c|c|c|c|c|c|c|c|c|c}
				\toprule \multirow{3}{*}{\textbf{T}} &
				\multirow{3}{*}{$\mathbf{\brho Y_{e}}$}  &\multicolumn{4}{|c|}{\textbf{$^{76}$Ge}} &
				\multicolumn{4}{c|}{\textbf{$^{76}$Se}}  &\multicolumn{4}{c}{\textbf{$^{82}$Se}}\\
				\cmidrule{3-6}  \cmidrule{7-10}  \cmidrule{11-14} & & 
				\multicolumn{2}{|c|}{\bf{R$_{i}$} [All]} &
				\multicolumn{2}{c|}{\bf{R$_{i}$} [U1F]} & 
				\multicolumn{2}{c|}{\bf{R$_{i}$} [All]} &
				\multicolumn{2}{c|}{\bf{R$_{i}$} [U1F]} & 
				\multicolumn{2}{c|}{\bf{R$_{i}$} [All]} &
				\multicolumn{2}{c}{\bf{R$_{i}$} [U1F]} \\ & & 
				\multicolumn{1}{c|}{\bf{$\lambda_{F} \ge \lambda_{BA}$}} &
				\multicolumn{1}{c|}{\bf{$\lambda_{BA}>\lambda_{F}$}} &
				\multicolumn{1}{c|}{\bf{$\lambda_{F} \ge \lambda_{BA}$}} &
				\multicolumn{1}{c|}{\bf{$\lambda_{BA}>\lambda_{F}$}}  & 
				\multicolumn{1}{c|}{\bf{$\lambda_{F} \ge \lambda_{BA}$}} &
				\multicolumn{1}{c|}{\bf{$\lambda_{BA}>\lambda_{F}$}} & 
				\multicolumn{1}{c|}{\bf{$\lambda_{F} \ge \lambda_{BA}$}} &
				\multicolumn{1}{c|}{\bf{$\lambda_{BA}>\lambda_{F}$}} &
				\multicolumn{1}{c|}{\bf{$\lambda_{F} \ge \lambda_{BA}$}}  & 
				\multicolumn{1}{c|}{\bf{$\lambda_{BA}>\lambda_{F}$}} &
				\multicolumn{1}{c|}{\bf{$\lambda_{F} \ge \lambda_{BA}$}}  & 
				\multicolumn{1}{c}{\bf{$\lambda_{BA}>\lambda_{F}$}} \\
				\hline
				1     & 10$^{4}$ & 77.6$^{(*)}$ &       &       & 1.6$^{(*)}$  & 1.0$^{(*)}$  &       &       & 1.0$^{(*)}$  & 66069.3$^{(*)}$ &       &       & 1.9$^{(*)}$ \\
				10    & 10$^{4}$ &       & 1.5  &       & 1.9  & 2.9  &       &       & 1.1  &       & 1.9  &       & 2.1 \\
				30    & 10$^{4}$ & 3.8  &       & 4.0 &       & 5.5  &       & 4.5  &       & 4.0  &       & 4.3  &     \\
				&       &       &       &       &       &       &       &       &       &       &       &       &  \\
				1     & 10$^{8}$ & 69.5$^{(*)}$ &       &       & 1.5$^{(*)}$  & 1.0$^{(*)}$  &       & 1.0$^{(*)}$  &       & 64416.9$^{(*)}$ &       &       & 1.9$^{(*)}$ \\
				10    & 10$^{8}$ &       & 1.5  &       & 1.9  & 2.9  &       &       & 1.1  &       & 1.9  &       & 2.1 \\
				30    & 10$^{8}$ & 3.8  &       & 4.0  &       & 5.5  &       & 4.5  &       & 4.0  &       & 4.3  &     \\
				&       &       &       &       &       &       &       &       &       &       &       &       &  \\
				1     & 10$^{11}$ & 1.0  &       & 1.0  &       & 1.0  &       & 1.0  &       & 1.0  &       & 1.0  &     \\
				10    & 10$^{11}$ & 1.4  &       & 1.4  &       & 1.5  &       & 1.4  &       & 1.4  &       & 1.4  &     \\
				30    & 10$^{11}$ & 7.7  &       & 6.5  &       & 8.8  &       & 6.7  &       & 8.2  &       & 7.2  &     \\
				\hline
				\multicolumn{2}{c|}{$\mathbf{\bar{R}}$} & \multicolumn{2}{c|}{18.7} & \multicolumn{2}{c|}{2.6} & \multicolumn{2}{c|}{3.3} & \multicolumn{2}{c|}{2.5} & \multicolumn{2}{c|}{14501.0} & \multicolumn{2}{c}{2.9} \\
				\hline
				\multicolumn{2}{c|}{$\mathbf{\bar{R}^{(*)}}$} &      \multicolumn{2}{c|}{3.0} & \multicolumn{2}{c|}{3.0} & \multicolumn{2}{c|}{4.0} & \multicolumn{2}{c|}{2.9} & \multicolumn{2}{c|}{3.2} & \multicolumn{2}{c}{3.2} 
	\end{tabular}}}
\end{table}
\begin{table}[h]
	\caption{\small R$_{i}$ [All] and R$_{i}$ [U1F] for three EC nuclei. See text for explanation of symbols.
	}\label{table:table13} \setlength{\tabcolsep}{2pt} 
	\centering {\small
		\resizebox{17cm}{!}{%
			\begin{tabular}{cc|c|c|c|c|c|c|c|c|c|c|c|c}
				\toprule \multirow{3}{*}{\textbf{T}} &
				\multirow{3}{*}{$\mathbf{\brho Y_{e}}$} & \multicolumn{4}{|c|}{\textbf{$^{88}$Sr}} &
				\multicolumn{4}{c|}{\textbf{$^{90}$Zr}}  &\multicolumn{4}{c}{\textbf{$^{128}$Te}}\\
				\cmidrule{3-6}  \cmidrule{7-10}  \cmidrule{11-14} & & 
				\multicolumn{2}{|c|}{\bf{R$_{i}$} [All]} &
				\multicolumn{2}{c|}{\bf{R$_{i}$} [U1F]} & 
				\multicolumn{2}{c|}{\bf{R$_{i}$} [All]} &
				\multicolumn{2}{c|}{\bf{R$_{i}$} [U1F]} & 
				\multicolumn{2}{c|}{\bf{R$_{i}$} [All]} &
				\multicolumn{2}{c}{\bf{R$_{i}$} [U1F]} \\ & & 
				\multicolumn{1}{c|}{\bf{$\lambda_{F} \ge \lambda_{BA}$}} &
				\multicolumn{1}{c|}{\bf{$\lambda_{BA}>\lambda_{F}$}} &
				\multicolumn{1}{c|}{\bf{$\lambda_{F} \ge \lambda_{BA}$}} &
				\multicolumn{1}{c|}{\bf{$\lambda_{BA}>\lambda_{F}$}}  & 
				\multicolumn{1}{c|}{\bf{$\lambda_{F} \ge \lambda_{BA}$}} &
				\multicolumn{1}{c|}{\bf{$\lambda_{BA}>\lambda_{F}$}} & 
				\multicolumn{1}{c|}{\bf{$\lambda_{F} \ge \lambda_{BA}$}} &
				\multicolumn{1}{c|}{\bf{$\lambda_{BA}>\lambda_{F}$}} &
				\multicolumn{1}{c|}{\bf{$\lambda_{F} \ge \lambda_{BA}$}}  & 
				\multicolumn{1}{c|}{\bf{$\lambda_{BA}>\lambda_{F}$}} &
				\multicolumn{1}{c|}{\bf{$\lambda_{F} \ge \lambda_{BA}$}}  & 
				\multicolumn{1}{c}{\bf{$\lambda_{BA}>\lambda_{F}$}} \\
				\hline
				1     & 10$^{4}$ & 1.9$\times$10$^{+09(*)}$ &       &       & 1.5$^{(*)}$  & 274789.4$^{(*)}$ &       &       & 1.0$^{(*)}$  & 1517.1$^{(*)}$ &       &       & 1.5$^{(*)}$ \\
				10    & 10$^{4}$ & 10.0 &       &       & 1.8  & 74.1 &       &       & 1.1  & 6.6  &       & 1.2  &     \\
				30    & 10$^{4}$ & 6.7  &       & 3.9  &       & 55.0 &       & 7.1  &       & 6.1  &       & 7.7  &     \\
				&       &       &       &       &       &       &       &       &       &       &       &       &  \\
				1     & 10$^{8}$ & 1.7$\times$10$^{+09(*)}$ &       &       & 1.5$^{(*)}$  & 1.1$^{(*)}$  &       & 1.0$^{(*)}$  &       & 95.1$^{(*)}$ &       &       & 1.3$^{(*)}$ \\
				10    & 10$^{8}$ & 9.9  &       &       & 1.8  & 70.6 &       &       & 1.1  & 6.6  &       & 1.2  &     \\
				30    & 10$^{8}$ & 6.7  &       & 3.9  &       & 55.0 &       & 7.1  &       & 6.1  &       & 7.7  &     \\
				&       &       &       &       &       &       &       &       &       &       &       &       &  \\
				1     & 10$^{11}$ & 1.0  &       & 1.0  &       & 1.0  &       & 1.0  &       & 1.0 &       & 1.0  &     \\
				10    & 10$^{11}$ & 1.8  &       & 1.3  &       & 5.4  &       & 1.3  &       & 3.0  &       & 2.9  &     \\
				30    & 10$^{11}$ & 9.8  &       & 5.3  &       & 53.8 &       & 9.3  &       & 12.4 &       & 12.6 &     \\
				\hline
				\multicolumn{2}{c|}{$\mathbf{\bar{R}}$} & \multicolumn{2}{c|}{4.0$\times$10$^{+08}$} & \multicolumn{2}{c|}{2.4} & \multicolumn{2}{c|}{30567.3} & \multicolumn{2}{c|}{3.3} & \multicolumn{2}{c|}{183.8} & \multicolumn{2}{c}{4.1} \\
				\hline
				\multicolumn{2}{c|}{$\mathbf{\bar{R}^{(*)}}$} &       \multicolumn{2}{c|}{6.6} & \multicolumn{2}{c|}{2.7} & \multicolumn{2}{c|}{45.0} & \multicolumn{2}{c|}{4.0} & \multicolumn{2}{c|}{6.0} & \multicolumn{2}{c}{4.9} 
	\end{tabular}}}
\end{table}
\begin{table}[h]
	\caption{\small R$_{i}$ [All] and R$_{i}$ [U1F] for three EC nuclei. See text for explanation of symbols. 
	}\label{table:table14} \setlength{\tabcolsep}{2pt} 
	\centering {\small
		\resizebox{17cm}{!}{%
			\begin{tabular}{cc|c|c|c|c|c|c|c|c|c|c|c|c}
				\toprule \multirow{3}{*}{\textbf{T}} &
				\multirow{3}{*}{$\mathbf{\brho Y_{e}}$} &\multicolumn{4}{|c|}{\textbf{$^{186}$Nd}}&
				\multicolumn{4}{c|}{\textbf{$^{195}$Tm}} &\multicolumn{4}{c}{\textbf{$^{204}$Pt}}\\
				\cmidrule{3-6}  \cmidrule{7-10}  \cmidrule{11-14} & & 
				\multicolumn{2}{|c|}{\bf{R$_{i}$} [All]} &
				\multicolumn{2}{c|}{\bf{R$_{i}$} [U1F]} & 
				\multicolumn{2}{c|}{\bf{R$_{i}$} [All]} &
				\multicolumn{2}{c|}{\bf{R$_{i}$} [U1F]} & 
				\multicolumn{2}{c|}{\bf{R$_{i}$} [All]} &
				\multicolumn{2}{c}{\bf{R$_{i}$} [U1F]} \\ & & 
				\multicolumn{1}{|c|}{\bf{$\lambda_{F} \ge \lambda_{BA}$}} &
				\multicolumn{1}{c|}{\bf{$\lambda_{BA}>\lambda_{F}$}} &
				\multicolumn{1}{c|}{\bf{$\lambda_{F} \ge \lambda_{BA}$}} &
				\multicolumn{1}{c|}{\bf{$\lambda_{BA}>\lambda_{F}$}}  & 
				\multicolumn{1}{c|}{\bf{$\lambda_{F} \ge \lambda_{BA}$}} &
				\multicolumn{1}{c|}{\bf{$\lambda_{BA}>\lambda_{F}$}} & 
				\multicolumn{1}{c|}{\bf{$\lambda_{F} \ge \lambda_{BA}$}} &
				\multicolumn{1}{c|}{\bf{$\lambda_{BA}>\lambda_{F}$}} &
				\multicolumn{1}{c|}{\bf{$\lambda_{F} \ge \lambda_{BA}$}}  & 
				\multicolumn{1}{c|}{\bf{$\lambda_{BA}>\lambda_{F}$}} &
				\multicolumn{1}{c|}{\bf{$\lambda_{F} \ge >\lambda_{BA}$}}  & 
				\multicolumn{1}{c}{\bf{$\lambda_{BA}>\lambda_{F}$}} \\
				\hline
				1     & 10$^{4}$ & ---     & ---     & ---     & ---     & 1.1$\times$10$^{+06(*)}$ &       &       & 12.0$^{(*)}$ & 31.8$^{(*)}$ &       &       & 8.5$^{(*)}$ \\
				10    & 10$^{4}$ & 55.7$^{(*)}$ &       &       & 5.1$^{(*)}$  &       & 1.3  & 1.2  &       &       & 1.6  &       & 3.5 \\
				30    & 10$^{4}$ & 14.9 &       & 2.6  &       & 15.4 &       & 12.1 &       & 3.3  &       & 3.7  &     \\
				&       &       &       &       &       &       &       &       &       &       &       &       &  \\
				1     & 10$^{8}$ & ---     & ---     & ---     & ---     & 1.1$\times$10$^{+06(*)}$ &       &       & 12.0$^{(*)}$ & 31.8$^{(*)}$ &       &       & 8.6$^{(*)}$ \\
				10    & 10$^{8}$ & 55.7$^{(*)}$ &       &       & 5.1$^{(*)}$  &       & 1.3  & 1.2  &       &       & 1.6  &       & 3.5 \\
				30    & 10$^{8}$ & 14.9 &       & 2.6  &       & 15.4 &       & 12.1 &       & 3.3  &       & 3.7  &     \\
				&       &       &       &       &       &       &       &       &       &       &       &       &  \\
				1     & 10$^{11}$ & 1.0$^{(*)}$  &       & 1.0$^{(*)}$  &       & 1.1  &       & 1.1  &       & 1.0  &       & 1.0  &     \\
				10    & 10$^{11}$ & 50.4 &       &       & 2.9  & 5.5  &       & 4.2  &       & 2.0  &       & 2.0  &     \\
				30    & 10$^{11}$ & 14.4 &       & 2.7  &       & 26.2 &       & 15.8 &       & 5.2  &       & 5.2  &     \\
				\hline
				\multicolumn{2}{c|}{$\mathbf{\bar{R}}$} & \multicolumn{2}{c|}{29.6} & \multicolumn{2}{c|}{3.2} & \multicolumn{2}{c|}{2.4$\times$10$^{+05}$} & \multicolumn{2}{c|}{8.0} & \multicolumn{2}{c|}{9.1} & \multicolumn{2}{c}{4.4} \\
				\hline
				\multicolumn{2}{c|}{$\mathbf{\bar{R}^{(*)}}$} &         \multicolumn{2}{c|}{23.6} & \multicolumn{2}{c|}{2.7} & \multicolumn{2}{c|}{9.5} & \multicolumn{2}{c|}{6.8} & \multicolumn{2}{c|}{2.6} & \multicolumn{2}{c}{3.2} 
	\end{tabular}}}
\end{table} 

\end{document}